# Biologically Inspired Predictive Coding TCN-Transformer for Anticipatory Human-Robot Interaction in Shared Physical Spaces

**Xiaoshan Zhou[1], Carol C. Menassa[2], and Vineet R. Kamat[3]**

[1]Ph.D. Candidate, Dept. of Civil and Environmental Engineering, University of Michigan, Ann Arbor, MI, 48109-2125. Email: xszhou@umich.edu

[2]Professor, Dept. of Civil and Environmental Engineering, University of Michigan, Ann Arbor, MI, 48109-2125. Email: menassa@umich.edu

[3]Professor, Dept. of Civil and Environmental Engineering, University of Michigan, Ann Arbor, MI, 48109-2125. Email: vkamat@umich.edu

**Abstract**

As mobile robots increasingly operate in environments shared with humans, proactively anticipating human motion rather than responding reactively is critical for preempting collisions during close-proximity navigation, while maintaining mobility efficiency and avoiding unnecessary yields. A timely and motivating engineering application is how autonomous vehicles interpret ambiguous right-of-way such as unsignalized pedestrian crossings. To address this challenge, this study explores the feasibility of decoding preparatory neural activity from wearable electroencephalography (EEG) to predict human motion intention before it is behaviorally expressed. Drawing inspiration from biological predictive coding mechanisms between the sensorimotor cortex and insula–frontoparietal network, we implement this principle in a Temporal Convolutional Network–Transformer architecture to decode fast-evolving EEG signals underlying perception–action transitions. In experiments involving twelve participants simulating road-crossing decisions under varying traffic volume, marked crosswalks, and traffic signals, neurophysiological analyses reveal hemispheric asymmetries in functional specialization and identify high-beta oscillations (16–25 Hz) in the right fronto-central region (F4) as robust neural markers of motor readiness and decision commitment. Using sliding-window feature extraction, we benchmarked sixteen classification models across traditional, recurrent, and convolutional deep learning architectures, and found that our approach achieved the highest Area Under the Curve (AUC) of 0.727 with an approximate 1-second look-ahead. This work demonstrates how biologically grounded temporal architectures can enhance anticipatory intelligence in autonomous systems and represents the first step toward proactive and adaptive human-robot interaction in the built environment.

## 1 Introduction

### 1.1 Motivating engineering problem

When mobile robots go beyond isolated within fenced workspaces to sharing unstructured physical environments with human users, this transition unlocks enormous potential for productivity and autonomy but also introduces significant safety challenges. The most immediate concern is the risk of human–robot collisions, particularly in dynamic settings (Markkula et al., 2020). Conventional robotic safety systems can detect obstacles and execute emergency stops, but such reactive mechanisms fail under occluded or ambiguous conditions.

A common example occurs when humans emerge unexpectedly from blind spots, such as behind shelves in warehouses, or behind parked vehicles and buses on urban roads (Frampton & Millington, 2022; Mole & Wilkie, 2017). In these scenarios, even the most advanced vision-based trajectory prediction models, whether using pose estimation (Kim et al., 2020; Lyu et al., 2024), body skeleton tracking (Jiang et al., 2022), or positional forecasting (Alahi et al., 2016; Saleh et al., 2017; Saleh et al., 2018), lose effectiveness because they depend on visible cues. To address these limitations, recent efforts in cooperative perception have explored Vehicle-to-Vehicle (V2V) and Vehicle-to-Infrastructure (V2I) sensing frameworks (Malik et al., 2023), where multiple agents share observations to maintain visibility of occluded targets (Koda et al., 2020; Tang et al., 2021). In addition, a complementary stream of research has focused on wearable and human-centered sensing, leveraging signals from GPS (Liebner et al., 2013), accelerometers (Flach et al., 2011), inertia sensors (Datta et al., 2014) in smartphones and wearable pressure sensors (Gao et al., 2016) to infer motion dynamics and walking patterns.

While such IoT-based cooperative sensing has extended environmental awareness, it largely remains reactive, i.e., tracking movement after it begins. However, the next step toward more robust and intelligent human–robot interaction requires a transition from passive detection to proactive, intent-aware perception. For instance, at a road intersection or in a warehouse, even if cooperative sensors predict a possible path overlap between a robot and a human, they cannot infer whether the human actually intends to cross. This uncertainty leads to inefficient behavior such as unnecessary braking or overly conservative yielding, which compromises both energy efficiency and traffic throughput.

This problem is especially acute in pedestrian safety, one of the most pressing challenges in intelligent transportation systems (Valos & Bennett, 2023). Pedestrians are unprotected traffic participants, and collision impacts often result in severe or fatal injuries (Anaya et al., 2014). According to the U.S. Governors Highway Safety Association, over 7,500 pedestrian fatalities occurred in 2022, the highest number in four decades (Kim, 2023). Particularly problematic are unsignalized crossings, where pedestrians' intentions are uncertain and right-of-way is ambiguous (Fu et al., 2019; Gerogiannis & Bode, 2024; Haleem et al., 2015; Noh et al., 2020; Olszewski et al., 2015). More importantly, these scenarios have been identified as the most safety-critical edge cases for autonomous vehicles (Brill et al., 2024; Nuñez Velasco et al., 2021). For a robotic vehicle or mobile platform, predicting whether a stationary pedestrian will cross or wait remains a central unsolved problem (Rasouli et al., 2018).

To bridge this perceptual gap from reactive sensing to proactive inference of internal decision states, this study explores the integration of wearable neurophysiological sensing through BCIs.

BCI technologies offer access to neural correlates of motor planning that precede physical motion, potentially providing robots with an anticipatory window into human intentions. By decoding preparatory EEG activity associated with motor readiness and perceptual decision-making, we can enable mobile robots and autonomous vehicles to anticipate pedestrians' crossing actions before movement onset, even when external cues are absent or ambiguous.

Accordingly, this research seeks to develop a computational framework for decoding brain dynamics underlying motor planning in road-crossing scenarios, establishing a proof of concept for cognition-aware human–robot interaction. Unlike prior BCI studies focused on explicit, cooperative tasks (e.g., handover (Cooper et al., 2020) or shared manipulation (Hu et al., 2024), this work targets implicit and uncertain interactions, where the human's decision is not externally cued but internally deliberated. Furthermore, while the readiness potential has long been studied as an event-related potential preceding self-initiated movement, it has primarily been examined in spontaneous or unconscious actions. However, our context demands identifying neural signatures of deliberate, reasoning-based intention formation. Establishing this connection between neurophysiological intent and real-world decision timing forms the foundation for the subsequent machine-learning framework for predicting motion initiation in human–robot shared spaces.

### 1.2 Electroencephalography (EEG) decoding

In this context, electroencephalography (EEG) and brain–computer interfaces (BCIs) represent a promising frontier. EEG provides direct access to neural dynamics underlying perception, cognition, and motor planning, capturing signals that precede physical movement (Donner et al., 2009; Parés-Pujolràs et al., 2023). Unlike vision-based skeleton tracking or inertial sensing via IMUs, both of which record motion after it begins, EEG data can reflect preparatory activity in the brain associated with intended actions before execution (Zhou & Liao, 2023).

Yet, most existing EEG decoding studies have focused on relatively static cognitive states, such as motor imagery classification (Ingolfsson et al., 2020; Lu et al., 2019), emotion recognition (He et al., 2022; Jia et al., 2020), or seizure detection (Zhang et al., 2018). These tasks evolve on a slow or discrete temporal scale and often lack the continuous temporal transitions characteristic of real-world decision processes. However, intent formation during perceptual decision-making, such as when a pedestrian evaluates whether to cross a street, unfolds as a dynamic and context-dependent process, involving continuous integration of sensory evidence, risk evaluation, and motor preparation (Schmidt & Färber, 2009; Sucha et al., 2017). Capturing such fast-evolving neural transitions thus requires decoding architectures capable of modeling fine-grained temporal dependencies in EEG signals.

Recent years have witnessed the emergence of deep learning architectures for EEG decoding, including convolutional neural networks (CNNs) (e.g., EEGNet (Lawhern et al., 2016), ShallowNet (Schirrmeister et al., 2017)), recurrent neural networks (RNNs) with Long-Short Term Memory (LSTM) (Kumar et al., 2021) or Gated Recurrent Unit (GRU) cells (Luo et al., 2018), and more recently, temporal convolutional networks (TCNs) (Walther et al., 2023) and Transformers (Wan et al., 2023). CNNs are effective in extracting spatial or spectral representations but remain limited in capturing long-range temporal dependencies. RNNs partially address this limitation but often suffer from vanishing gradients and limited parallelization. TCNs introduced causal and dilated convolutions, enabling efficient modeling of long-term dependencies with stable gradients, while maintaining temporal causality, a desirable property for real-time intent prediction.

To further enhance temporal representation, researchers have introduced attention mechanisms as auxiliary modules to focus on the most informative features, including spatial (Squeeze-Excitation) (Zhang et al., 2024), temporal (self-attention) (Huang et al., 2024), or cross-channel (multi-head attention) (Zhou et al., 2025). These mechanisms echo biological attention processes in the human brain, selectively amplifying salient information while suppressing irrelevant inputs. The cross-fertilization between neuroscientific principles and artificial intelligence (AI) model design has thus become mutually reinforcing: brain-inspired computational models improve EEG decoding, and in turn, these models offer new tools to interpret neural dynamics.

### 1.3 Predictive Coding

Building on this foundation, predictive coding has emerged as one of the most influential frameworks unifying brain function and intelligent behavior. Rooted in the free energy principle and active inference, it conceptualizes the brain as a hierarchical system that continuously predicts incoming sensory input and minimizes prediction errors through feedback correction (Friston & Kiebel, 2009). Rather than being a passive receiver of information, the brain operates as an active inferential organ: constantly generating hypotheses about the external world and updating them when reality deviates from expectation (Sinclair et al., 2021). This dynamic interplay between prediction and feedback allows for efficient information processing and adaptive behavior in uncertain environments.

From a neurophysiological standpoint, this mechanism is implemented through a distributed predictive hierarchy involving cortical and subcortical loops. Higher associative regions such as the hippocampus, insula, and cerebellum generate internal models of sensory and motor contingencies, drawing from prior experience to anticipate upcoming inputs (Craig & Witton, 2022). Prediction errors, arising from mismatches between expected and actual sensory feedback, are propagated upward from lower sensory–motor cortices to update these internal models (Grob et al., 2024). Subcortical structures, including the periaqueductal gray (PAG), superior colliculus, and amygdala, play a crucial role in mediating rapid defensive and autonomic responses (Reis et al., 2023), while cortical regions such as the frontal–parietal network arbitrate among competing predictions to guide adaptive action selection. The cerebellum–hippocampus–insula axis functions as an integrative hub—receiving multisensory feedback, computing prediction errors, and redistributing attentional and mnemonic resources to maintain coherence between perception, memory, and action (Huang et al., 2021; Ito, 2008; Menon & Uddin, 2010). This continuous calibration of expectations and outcomes forms the physiological substrate of anticipatory cognition, enabling organisms to act proactively rather than reactively.

Inspired by these insights, the present study introduces a Predictive-Coding TCN–Transformer architecture, designed to emulate this hierarchical feedback system in a computational model of intent decoding. The proposed architecture integrates feedforward evidence accumulation (implemented via the TCN to capture short-horizon, causal dynamics and the Transformer to capture long-range, non-local dependencies) with feedback-based predictive correction (realized by a predictive-coding recurrence that generates top-down predictions and minimizes the ensuing prediction errors to update lower-level representations), reflecting the bidirectional information flow observed in fronto-parietal and cerebellar–cortical circuits. By doing so, this approach not only advances EEG-based intent prediction for dynamic perceptual decision-making but also provides a biologically grounded framework for exploring how hierarchical temporal models can reproduce the anticipatory, self-correcting nature of human cognition.

In summary, the main contributions of this paper are as follows:

1. Problem formulation: We introduce EEG-based intent prediction as a critical enabler for safe and anticipatory human–robot interaction in shared physical spaces. Moreover, rather than focusing on abstract decoding tasks, we ground the study in a representative and safety-critical scenario, pedestrian road crossing, that captures the temporal evolution in perceptual–cognitive–motor decision-making under real-world constraints. This case study exemplifies the broader class of intent-aware robotics applications in intelligent transportation systems.
2. Neurophysiological grounding: We conduct a detailed analysis of EEG signals to identify neural features that mediate perceptual decision-making and motor preparation. Through time–frequency decomposition and spatial–spectral analysis, we isolate high-beta (16–25 Hz) oscillations in the right fronto-central region (F4) as a key correlate of preparatory activity leading to movement initiation. This finding aligns with prior literature on motor readiness potentials and fronto-parietal coupling during action planning, providing a neurophysiological foundation for feature selection and interpretability.
3. Biologically-inspired algorithm: We propose a Predictive-Coding TCN–Transformer–TCN framework that transforms the network from a purely feedforward sequence model into a bi-directional inference system that: (1) preserves sensory causality and precision (TCN, bottom-up encoding), (2) learns hierarchical temporal abstractions (Transformer, top-down generative context), and (3) iteratively reconciles both through error minimization (predictive coding loop).
4. Empirical validation: We collected a new EEG dataset involving twelve participants performing a mental road-crossing decision task, simulating the cognitive transition from perception to action readiness. Each participant completed five trials of mentally evaluating road-crossing opportunities under controlled visual stimuli. The EEG signals were preprocessed, segmented with sliding windows, and labeled based on decision phases. The final dataset comprises hundreds of temporally structured samples depending on window length and stride (e.g., 838 segments for 9×3 configuration). Using this dataset, we benchmarked sixteen classifiers, including conventional machine learning baselines, RNNs, and TCN variants, under standardized cross-validation protocols.

5. Interpretive insight: Experimental results demonstrate that models employing causal temporal convolutions consistently outperform recurrent and static baselines in short transient windows. The proposed predictive-coding TCN–Transformer achieves the highest area under the ROC curve. Moreover, ablation experiments with varying feature channels and window parameters show that performance gains are not dataset artifacts but reflect principled modeling of neurocognitive dynamics. These findings contribute both to the design of biologically inspired temporal networks and to the understanding of how anticipatory brain signals can inform intent-aware robotic systems.

The remainder of this paper is organized as follows. Section 2 reviews related works on TCN architectures for EEG decoding. Section 3 details the proposed methodology and modeling framework. Section 4 presents results and comparative evaluations. Section 5 discusses implications, limitations, and future directions.

## 2 Related work

Deep learning has become the dominant approach for decoding EEG signals in recent years, largely due to its capacity to model nonlinear dynamics and multi-scale temporal dependencies. Among these methods, the TCN has emerged as a particularly effective architecture for processing sequential neural data because of its causal structure, large receptive field, and stable gradient propagation.

The earliest applications of TCN to EEG analysis focused on clinical diagnosis tasks such as epileptic seizure detection (Zhang et al., 2018), followed by motor imagery classification ((Ingolfsson et al., 2020; Lu et al., 2019) and emotion recognition (He et al., 2022; Jia et al., 2020). These studies demonstrated that temporal convolutions could outperform both convolutional neural networks (CNNs), which emphasize spatial feature extraction (Ingolfsson et al., 2020; Musallam et al., 2021), and recurrent architectures such as LSTM and GRU, which often suffer from vanishing gradients and limited parallelization (Lu et al., 2020; Wang et al., 2022). A systematic benchmark by (Walther et al., 2023) quantified this advantage, reporting an 8.6% average accuracy improvement of TCNs over recurrent networks across multiple EEG benchmarks, establishing TCNs as a new state-of-the-art backbone for temporal EEG decoding.

With growing recognition that EEG signals are highly redundant and channel-dependent, feature selection became a key research direction. For example, (Wang et al., 2025) employed a Fisher-score channel selection method before feeding data into a TCN for seizure detection, enhancing signal-to-noise ratio and reducing model overfitting. Subsequent studies introduced attention mechanisms to further improve feature weighting and dynamic temporal adaptation. The landmark ATCNet (Altaheri et al., 2023) combined TCN layers with a self-attention block, achieving state-of-the-art results for motor imagery decoding.

Later works extended this paradigm by incorporating specialized attention modules: multi-anchor attention for adaptive temporal dependency learning (Ding et al., 2024); self-attention layers for epilepsy detection (Huang et al., 2024); multi-head attention for motor imagery (Chunduri et al., 2024; Liang et al., 2024; Liu et al., 2025; Xie et al., 2024; Zhao et al., 2025); channel-wise attention (Qin et al., 2024); and Squeeze-Excite-Compress modules that dynamically reweight feature maps (Zhang et al., 2024). (Yang et al., 2024) further integrated factorization machines and separable convolutions into TCNs for modeling cross-feature interactions in emotion recognition. (Lu et al., 2024) and (Xu et al., 2023) extended attention-TCN frameworks to EEG reconstruction and sleep-stage classification, respectively, confirming the versatility of temporal convolutional modeling.

Parallel research explored hybrid architectures that couple TCNs with other neural operators to capture complementary feature hierarchies. For instance, (Zhang et al., 2019) concatenated CNN and TCN outputs for mental-workload assessment, while (Bao et al., 2024) extracted channel-wise spectral features via CNNs before feeding them into a TCN for sleep-stage classification. Several studies fused TCNs with recurrent units, such as TCN-LSTM models for seizure detection (Dong et al., 2024; Sun et al., 2023). Others explored CNN–TCN or Graph Neural Network (GNN)–TCN frameworks to enrich spatial–temporal representations across electrodes, for instance, CNN–TCN for depression scoring (Hashempour et al., 2022), schizophrenia detection (Sharma et al., 2023), and emotion recognition (Wu et al., 2024); and GNN–TCN architectures for motor imagery, seizure detection and emotion recognition (Ji et al., 2024; Li et al., 2024; Shi et al., 2024).

Further innovations combined TCNs with Generative Adversarial Networks (GANs) for data augmentation (Hu et al., 2024) and bidirectional GRUs for capturing forward-backward context (Zhu et al., 2024). (Qian et al., 2024) proposed a convolutional-recurrent front-end feeding into a TCN with multi-head attention, yielding strong performance in EEG-based emotion recognition. Collectively, these efforts established that integrating TCNs with complementary temporal or spatial mechanisms yields consistent performance gains across diverse EEG tasks.

More recent studies have focused on combining Transformers with TCNs to merge global attention and local causal modeling. (Su et al., 2024) introduced a TCN–Transformer hybrid where the TCN extracted preliminary spectral–spatial features that were refined by a Transformer encoder, achieving state-of-the-art emotion recognition. In addition, (Pesen & Altuncu, 2025) proposed a Transformer–TCN structure where Transformer layers captured global spatial relationships before TCN modules modeled temporal dependencies. Shi et al. (2024b) advanced this direction by proposing a dual-branch Transformer–TCN architecture trained jointly on convolutional embeddings. (Altaheri et al., 2025) further designed a dual-headed Transformer with two TCN branches for EEG-based motor imagery decoding, achieving top accuracy on the BCI Competition IV datasets.

These studies collectively confirm that hybridizing attention-based Transformers with TCNs leverages the strengths of both: Transformers provide long-range dependency modeling, while TCNs preserve causality and local receptive efficiency. Indeed, (Chaudhary et al., 2024) demonstrated that Transformer-based networks can surpass pure TCNs on fine-grained motor-imagery decoding, motivating a growing trend toward attention-convolution fusion. (Zhao et al., 2026) further integrated TCN layers into a Transformer backbone, improving accuracy by 2% over the strongest baseline on the same dataset.

Despite these advancements, existing architectures primarily focus on classification rather than prediction, emphasizing pattern recognition over temporal inference of evolving cognitive processes. This gap is especially critical for neurophysiological sensing applications in dynamic environments, such as a human's decision to cross in front of an approaching robotic vehicle, where risk assessment and intention emerge within milliseconds and govern underlying neural activities. Moreover, although attention mechanisms mimic selective weighting in human perception, they lack a top-down predictive component that characterizes biological information processing, which is however crucial in humans' predictive coding mechanism for danger awareness. Our work therefore implements predictive-Coding algorithmically within a TCN–Transformer–TCN framework, embedding feedback loops that generate anticipatory internal predictions and correct them via precision-weighted error signals. This biologically inspired mechanism aims to enable more stable and context-aware temporal learning, aligning algorithmic inference with the hierarchical predictive processes observed in cortical dynamics.

## 3 Methodology

### 3.1 EEG dataset

W collected a new EEG dataset that captures the perceptual–cognitive–motor temporal continuum underlying the decision of when to act (i.e., when to initiate crossing). Twelve healthy volunteers (six males and six females, mean age = 24.92 years) were recruited for the experiment. The chosen sample size was consistent with contemporary EEG research norms: a recent biometric review (Cheng et al., 2022) reported that EEG studies in civil engineering typically recruit an average of twelve participants. Furthermore, the sufficiency of the sample was validated empirically by the convergence behavior of the Hidden Markov Model (HMM) used in subsequent modeling stages, confirming stable parameter estimation across participants.

For all statistical analyses, data normality was verified before hypothesis testing, and Cohen's d = 0.4 was used as a benchmark for the minimal acceptable effect size, consistent with psychological and cognitive neuroscience standards (Brysbaert, 2019). All participants reported normal or corrected-to-normal vision, with no history of neurological disorders such as epilepsy or brain injury. Informed consent was obtained verbally prior to participation, and the study protocol was approved by the University of Michigan Institutional Review Board (Reference ID: HUM00249262).

Participants received a preparatory instruction email one day before the experiment, advising them to obtain sufficient rest, refrain from consuming caffeinated beverages, and wash their hair with mild shampoo to optimize electrode contact. The sessions were conducted in a dimly lit, sound-attenuated environment to minimize visual and auditory distractions. Before EEG recording, participants completed a brief demographic questionnaire (age, gender, education), after which the experimenter assisted with headset placement.

The EEG was recorded using a 14-channel EPOC X headset, with electrodes positioned according to the international 10–20 system (Klem et al., 1999). Saline-soaked felt pads were used to ensure stable contact, and electrode impedance was continuously monitored through EmotivPro software (EMOTIV, 2023) until all sensors achieved optimal contact quality. EEG data were sampled at 128 Hz with an online 0.16 Hz first-order high-pass filter applied to remove DC offset.

Each trial began with a 500 ms fixation cross ("+"), displayed at the center of the screen to stabilize visual attention and minimize eye-movement artifacts. This was followed by a short animated video clip depicting a road-crossing scene from a pedestrian perspective. Participants were instructed to imagine themselves standing at the curb, observing traffic, and to press the "up" key when they judged it was safe to cross. Each animation looped continuously until a response was made, ensuring self-paced decision-making without temporal constraints. Upon response, the next scenario was automatically initiated.

The experimental design incorporated five distinct traffic scenarios, each simulating real-world intersection conditions with varying degrees of complexity, as summarized in Table 1 and illustrated in Figure 1. Scenario parameters were informed by prior literature on pedestrian safety and behavioral modeling (Fu et al., 2019; Gerogiannis & Bode, 2024; Haleem et al., 2015; Noh et al., 2020; Olszewski et al., 2015). Approximately 30% of pedestrian accidents occur at non-signalized crosswalks, highlighting the importance of including such conditions. Traffic density classifications were based on (Homburger et al., 1982), ensuring ecological validity across low-, medium-, and high-density road conditions. Additional variations, such as right-turning vehicles (Kumar et al., 2019) and intersection geometry ((Kim et al., 2020; Koehler et al., 2013; Volz et al., 2016), were incorporated to enhance the diversity of perceptual and decision-making demands.

Table 1. Descriptions of the simulated experimental scenarios.

| **Traffic Scenarios** | **Description** |
|---|---|
| Minimal Traffic Volume | A two-lane road without signalized crosswalks, with no cars approaching within sight. |
| Low Traffic Volume | A two-lane road without signalized crosswalks, with two cars approaching at a slow speed within sight. |
| High Traffic Volume | A two-lane road without signalized crosswalks, with two cars approaching in the adjacent lane and three vehicles approaching in the opposite lane at a moderate speed within sight. |
| Surface-marked Intersection | A four-way intersection with surface-marked crosswalks but without traffic lights. One car is waiting behind the sidewalk on the side opposite to the pedestrian's intended movement, with its right-turn signal flashing and visible within sight. |
| Signalized Intersection | A four-way intersection with surface-marked crosswalks and a traffic light (green/red). One car is waiting behind the sidewalk in the vertical orthogonal direction relative to the pedestrian's intended movement with its right-turn signal flashing. A red traffic light, perpendicular to the pedestrian's movement, is visible at the intersection. |

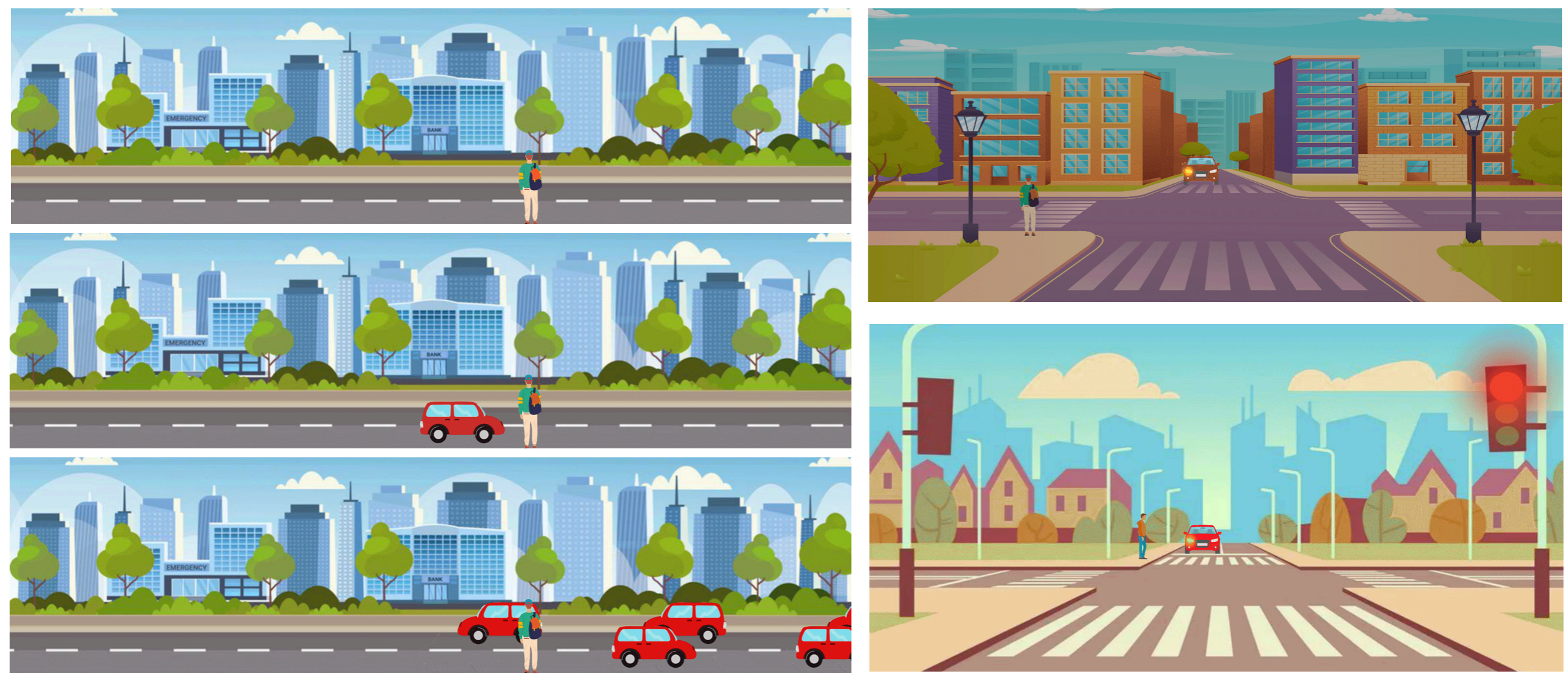

Figure 1. Experimental Stimuli (presented as animations to participants).

All stimuli were designed in Adobe Animate. The animations were created using Motion Tween (Adobe, 2023) at a frame rate of 24 FPS to produce realistic vehicular motion. Visual cues such as traffic lights, turn signals, and ambient lighting were rendered with keyframe-based flicker effects and glow filters to approximate real-world luminance conditions. Stimuli were presented using PsychoPy v2022.2.5 (Peirce et al., 2019), and the corresponding GIF versions are provided in the supplementary materials.

### 3.3 EEG pre-processing

EEG recordings represent temporal voltage fluctuations measured across multiple scalp electrodes, capturing the brain's electrophysiological activity in the time domain. To extract meaningful neural signatures from these raw signals, we applied a structured pre-processing pipeline comprising three analytical dimensions: (i) spectral decomposition via Power Spectral Density (PSD), (ii) time–frequency representation (TFR) to capture dynamic oscillatory patterns, and (iii) functional connectivity analysis to assess inter-regional synchronization. This multi-layered approach aims to provide a comprehensive understanding of both localized and distributed neural processes involved in this perceptual–motor decision-making.

#### 3.3.1 Power Spectral Density (PSD)

To decompose the EEG signals into their constituent frequency components, the Fast Fourier Transform (FFT) (Cochran et al., 1967) was applied to each channel. For a discrete signal $x(t)$ sampled at frequency $f_s$, the FFT computes the complex spectrum:

$$X(f_s) = \sum_{t=0}^{N-1} x(t) e^{-j2\pi f_k t/N}, \quad f_k = \frac{k f_s}{N}, \qquad k = 0,1, \dots, N-1 \qquad (1)$$

where $X(f_k)$ contains both magnitude and phase information for each frequency bin $f_k$.

To mitigate spectral leakage caused by discontinuities at signal boundaries, a Hanning-tapered window was applied before the FFT, preventing artificial energy spread across frequencies.

The PSD quantifies the distribution of power per unit frequency, computed as:

$$P(f_k) = \frac{|X(f_k)|^2}{\Delta f} \quad (2)$$

where $\Delta f = f_s/N$ is the frequency resolution. The PSD was estimated over five canonical EEG bands: theta (4–8 Hz), alpha (8–12 Hz), low beta (12–16 Hz), high beta (16–25 Hz), and gamma (25–45 Hz). Given 14 electrodes, the resulting PSD feature matrix had 70 spectral dimensions continuously sampled at 8 Hz.

Epoch segmentation was performed using synchronized event markers corresponding to stimulus onset and participant response, ensuring alignment between EEG recordings (EmotivPRO) and stimulus presentation (PsychoPy). To achieve precise synchronization, PsychoPy was programmed to send event triggers directly to the EEG system, embedding a shared reference timestamp for each trial.

While PSD effectively summarizes the global spectral power distribution, it provides only an average representation over an epoch and does not retain temporal evolution. To capture dynamic changes in oscillatory activity, time–frequency analysis was subsequently applied.

### 3.3.2 Time–Frequency Representation (TFR)

To analyze how oscillatory power evolves over time, we computed the TFR of EEG signals using two complementary methods:

1. Morlet Wavelet Transform –

The EEG signal $x(t)$ was convolved with a complex Morlet wavelet:

$$\psi_f(t) = e^{i2\pi ft} e^{-t^2/(2\sigma_t^2)} \quad (3)$$

where $f$ is the center frequency and $\sigma_t$ determines temporal resolution.

This method provides high temporal precision but relatively coarse frequency resolution, suitable for capturing transient cognitive events.

2. Multitaper Spectral Estimation –

Following (Slepian, 1978), multiple orthogonal tapering windows $h_m(t)$ were applied to the signal to compute independent spectral estimates:

$$P(f) = \frac{1}{M} \sum_{m=1}^{M} \left| \int x(t) h_m(t) e^{-i2\pi ft} dt \right|^2 \quad (4)$$

where $M$ is the number of tapers. Averaging across tapers reduces variance and enhances spectral stability, achieving higher frequency precision albeit with coarser temporal resolution.

Both analyses were conducted in MNE-Python (Gramfort et al., 2013) after exporting EEG signals in European Data Format (.edf) from EmotivPRO. To visualize event-related changes in spectral power, time–frequency maps were baseline-normalized using a 200 ms pre-stimulus interval. Two normalization schemes were applied:

- Arithmetic mean correction: subtracting the baseline mean power $\bar{P}_{base}$ from each time–frequency point.
- Log-ratio transformation:

$$P_{norm}(t,f) = \log\left(\frac{P(t,f)}{\bar{P}_{base}(f)}\right) \tag{5}$$

which provides a scale-invariant measure of relative power change and improves interpretability across participants.

These transformations yielded temporally resolved maps of power modulations across EEG frequency bands, revealing how neural oscillations evolved from perception to motor preparation during crossing decisions.

### 3.3.3 Functional Connectivity

While activation-based analyses identify localized power fluctuations, effective cognitive processing arises from coordinated interactions across distributed brain regions. To quantify such inter-regional synchronization, we computed the Weighted Phase Lag Index (wPLI) (Vinck et al., 2011), a robust connectivity metric that mitigates spurious correlations caused by volume conduction.

For two EEG signals $i$ and $j$, let $S_{ij}(f)$ denote their complex cross-spectrum, defined as:

$$S_{ij}(f) = X_i(f)X_j^*(f) \tag{6}$$

where $X_j^*(f)$ is the complex conjugate of $X_j(f)$. The wPLI measures the consistency of the imaginary part of the cross-spectrum across trials:

$$wPLI_{ij}(f) = \frac{\left|\mathrm{E}[Im(S_{ij}(f))]\right|}{\mathrm{E}[\left|Im(S_{ij}(f)\right|]} \tag{7}$$

where $Im(\cdot)$ denotes the imaginary component, and $\mathrm{E}[\cdot]$ indicates averaging across time or trials.

By focusing exclusively on non-zero phase lags, wPLI eliminates the influence of common-source artifacts and volume conduction effects that often bias traditional metrics such as coherence or phase-locking value (PLV).

Values of wPLI range from 0 (no consistent lagged synchronization) to 1 (perfect phase-lag consistency), reflecting the strength of directional, time-delayed coupling between neural sources.

Through this connectivity analysis, the study captures not only the local spectral activations but also the network-level coordination underpinning perceptual evaluation and motor-intention formation. Together, the PSD, TFR, and wPLI analyses aims to provide a multiscale depiction of how the brain dynamically transitions from perception to decision and action, forming the neural substrate for predictive modeling of crossing intention.

## 3.4 Proposed Predictive-Coding TCN-Transformer framework

### 3.4.1 Overview

As shown in Figure 2, the proposed framework, termed the Predictive-Coding TCN-Transformer, integrates three core modules—temporal convolutional encoding, Transformer-based global context modeling, and top-down predictive feedback—into a unified predictive coding loop. The architecture is implemented as a hierarchical encoder–decoder structure, where local temporal features are first extracted by a causal TCN. These features represent short-term dependencies within EEG sequences, analogous to the localized spatiotemporal processing observed in cortical columns of the brain's sensorimotor areas.

The encoded features are then projected into a Transformer encoder, which captures long-range dependencies and contextual relationships among temporally distant EEG patterns. The Transformer serves as a high-level inference module functionally similar to the brain's associative networks that generates top-down predictions of expected local representations. These predicted representations are compared with the original TCN activations to compute prediction errors, which in turn drive feedback updates to refine the local feature representation iteratively.

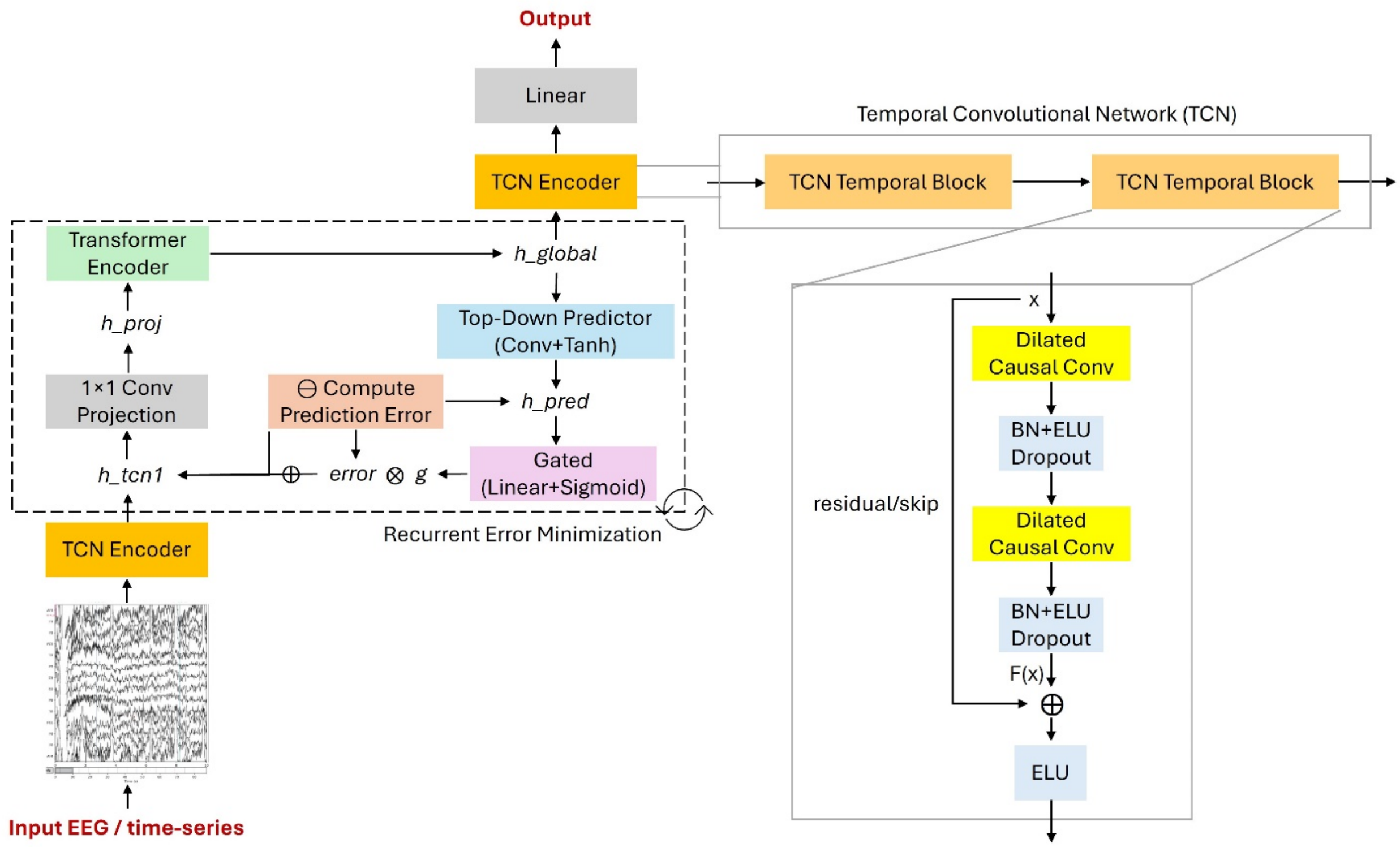

Figure 2. Overview of the proposed Predictive-Coding TCN–Transformer framework.

Mathematically, this predictive loop can be formalized as:

$$h_t^{(l+1)} = h_t^{(l)} + g^{(l)} \odot (\hat{h}_t^{(l)} - h_t^{(l)}) \tag{8}$$

where $h_t^{(l)}$ denotes the local feature representation at layer $l$, $\hat{h}_t^{(l)}$ is the Transformer-predicted feature at the same level, and $g^{(l)} \in [0, 1]$ is an adaptive feedback gain learned through a gating mechanism that regulates the influence of top-down corrections. This iterative refinement mechanism embodies the principle of free-energy minimization in predictive coding, i.e., reducing the discrepancy between internal predictions and sensory evidence across successive inference cycles.

After several feedback iterations, the refined high-level features are decoded through a secondary TCN, which restores temporal resolution and produces the final intention likelihood output through a fully connected layer. This design allows the model to concurrently leverage:

- Local precision (via causal convolutions capturing fast sensorimotor fluctuations),
- Global context (via self-attention for sequence-wide dependencies), and
- Bi-directional inference (via top-down feedback for error correction).

### 3.4.2 Temporal Convolutional Network (TCN)

The TCN module transforms the fused temporal features into a representation that is causal, multi-scale, and stable for downstream intention decoding. It is implemented as a short stack of residual temporal blocks with dilated causal convolutions, batch normalization (BN), ELU activations, and dropout. Causality ensures that predictions at time $t$ depend only on samples $\leq t$, which is critical for real-time BCI and intent anticipation.

In our architecture we use two TCN stages: $TCN_1$ (an encoder operating close to the input) and $TCN_2$ (a refinement head after predictive feedback). The formulation below applies to both; for concreteness we detail $TCN_1$.

Let $x \in \mathbb{R}^{B \times C_{in} \times T}$ denote a mini-batch of sequences (batch $B$, channels $C_{in}$, time $T$). For $TCN_1$, we set $C_{in} = 1$ (a per-time feature after preliminary mixing).

A 1-D dilated causal convolution with kernel size $k$ and dilation $d$ at time index $t$ is

$$y_c^{(t)} = \sum_{i=0}^{k-1} \sum_{c'} w_{c,c',i} x_{c'}^{(t-i \cdot d)} \quad for\ t \geq 0 \tag{9}$$

and undefined for $t - i \cdot d < 0$ (feature leakage is prevented). In practice we use left padding $p = (k-1)d$ and (optionally) a "chomp" layer to maintain length $T$ while preserving strict causality. The reason for using dilation is that increasing $d$ spreads the kernel taps over longer horizons, giving a large effective receptive field without extra parameters or loss of resolution.

Each TemporalBlock contains two dilated causal conv layers at the same dilation $d$, with BN–ELU–Dropout after each conv, and a  path:

$$z_1 = Drop(ELU(BN(Conv_{k,d}(x)))) \tag{10}$$

$$z_2 = Drop(ELU(BN(Conv_{k,d}(z_1)))) \tag{11}$$

If the input/output channel counts differ, the skip uses a 1×1 conv: $s = W_s * x; otherwise\ s = x$. The block output is

$$y = ELU(z_2 + s) \tag{12}$$

Residual connections preserve low-frequency trends and stabilize gradients; ELU reduces dead activations relative to ReLU and works well with BN for non-stationary EEG.

We use two residual blocks ($L = 2$) with exponentially increasing dilation:

$$d_l = 2^l, \quad l = 0,1 \tag{13}$$

The channel schedule is 1→32→64. Concretely:

- Block 1: $C_{in} = 1$, $C_{out} = 32$, $k = 3$, $d = 1$, padding $p = 2$
- Block 2: $C_{in} = 32$, $C_{out} = 64$, $k = 3$, $d = 2$, padding $p = 4$

Dropout is set to 0.2 after each conv. Stride is 1 everywhere, so the temporal length is preserved.

For a block with two convs (both kernel $k$ and dilation $d$), the increment to the receptive field equals

$$\Delta RF_{block} = 2(k-1)d \tag{14}$$

Stacking $L$ blocks with $d_l = 2^l$ yields the closed form

$$RF(L,k) = 1 + 2(k-1)\sum_{l=0}^{L-1} 2^l = 1 + 2(k-1)(2^L - 1) \tag{15}$$

For our settings ($L = 2$, $k = 3$),

$$RF = 1 + 2 \cdot (3-1)\,(2^2 - 1) = 13 \; time\ steps \tag{16}$$

An illustration of the TCN layer processing is provided in Figure 3. Thus, each output at time $t$ aggregates information from $[t-12, t]$. If EEG is sampled at 256 Hz (128 Hz), this corresponds to ≈50ms (≈100 ms). Extending to a third block with $d = 4$ would raise RF to 29 steps (≈113ms at 256 Hz; ≈226 ms at 128 Hz).

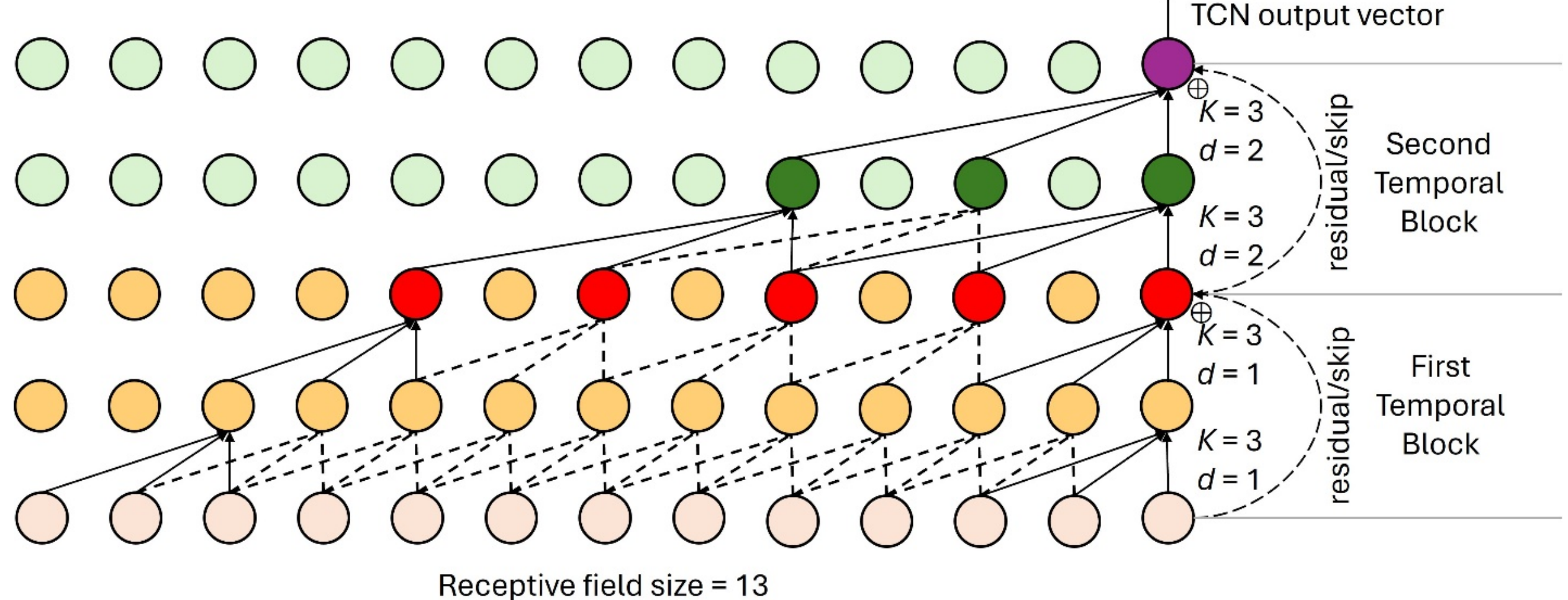

Figure 3. Architecture of the TCN used in this study. Each temporal block contains two dilated causal convolution layers with kernel size *K*=3 and dilation factors *d=1* (first block) and *d*=2 (second block).

Let $T$ be the input length: after Block 1: $[B, 1, T] \to [B, 32, T]$; after Block 2: $[B, 32, T] \to [B, 64, T]$. Length $T$ is unchanged due to padding $p = (k-1)d$ and stride 1. The TCN head (TCN$_2$) used after feedback has the same form but takes $d_{model}$ channels as input and outputs a refined temporal representation; we use the last time step as the summary statistic for the classifier:

$$h_{TCN} \in \mathbb{R}^{B \times C_{out} \times T} \Rightarrow \hat{y} = W_{fc} h_{TCN}[:,:,T-1] + b \quad (17)$$

Moreover, padding $p = (k-1)d$ on the left ensures that the earliest valid index of the convolution aligns with time $t$, preventing access to future samples $t+i$ (strict causality). Because stride = 1 and we use residual connections, the model is information non-losing over its RF: features are transformed but temporal alignment is preserved, which is important for fusing with attention-based context later.

In terms of computational complexity, for a Conv1D with $C_{in}$, $C_{out}$, $k$, $t$, the MACs are $O(C_{in}\ C_{out}\ k\ t)$; dilation does not change compute, only the spacing of taps. Our two-block encoder therefore scales linearly with sequence length and is markedly cheaper than attention over long windows, making it suitable for edge or online deployment.

### 3.4.3 Transformer Encoder

Whereas the preceding TCN encoder (TCN$_1$) extracts local, short-range temporal features while preserving causality, the Transformer encoder is introduced to capture long-range dependencies and global contextual interactions across the entire temporal window.

We employ the *Transformer encoder layer* as implemented in PyTorch, parameterized as follows (Table 2):

Table 2. Parameters in the Transformer encoder.

| Parameter | Description | Value |
|---|---|---|
| $d_{model}$ | Embedding dimension | 64 |
| $n_{head}$ | Number of attention heads | 2 |
| $d_{ff}$ | Feed-forward layer width | (4 $d_{model}$= 256) |
| Dropout | Regularization rate | 0.2 |
| $N_L$ | Number of stacked layers | 1 |

Each encoder layer is composed of two submodules: 1) Multi-Head Self-Attention (MHSA) (Vaswani et al., 2017), and 2) Position-wise Feed-Forward Network (FFN).

These are combined with residual connections, layer normalization, and dropout, forming the following forward equations:

$$Z_1 = LN(X + Drop(MHSA(X))) \quad (18)$$

$$Y = LN(Z_1 + Drop(FFN(Z_1))) \quad (19)$$

where $x \in \mathbb{R}^{B \times T \times d_{model}}$ is the sequence of projected features output by TCN$_1$.

The MHSA mechanism allows each time step to attend to all other steps, enabling the encoder to infer temporal dependencies beyond the receptive field of the TCN. A schematic of the Transformer mechanism is shown in Figure 4.

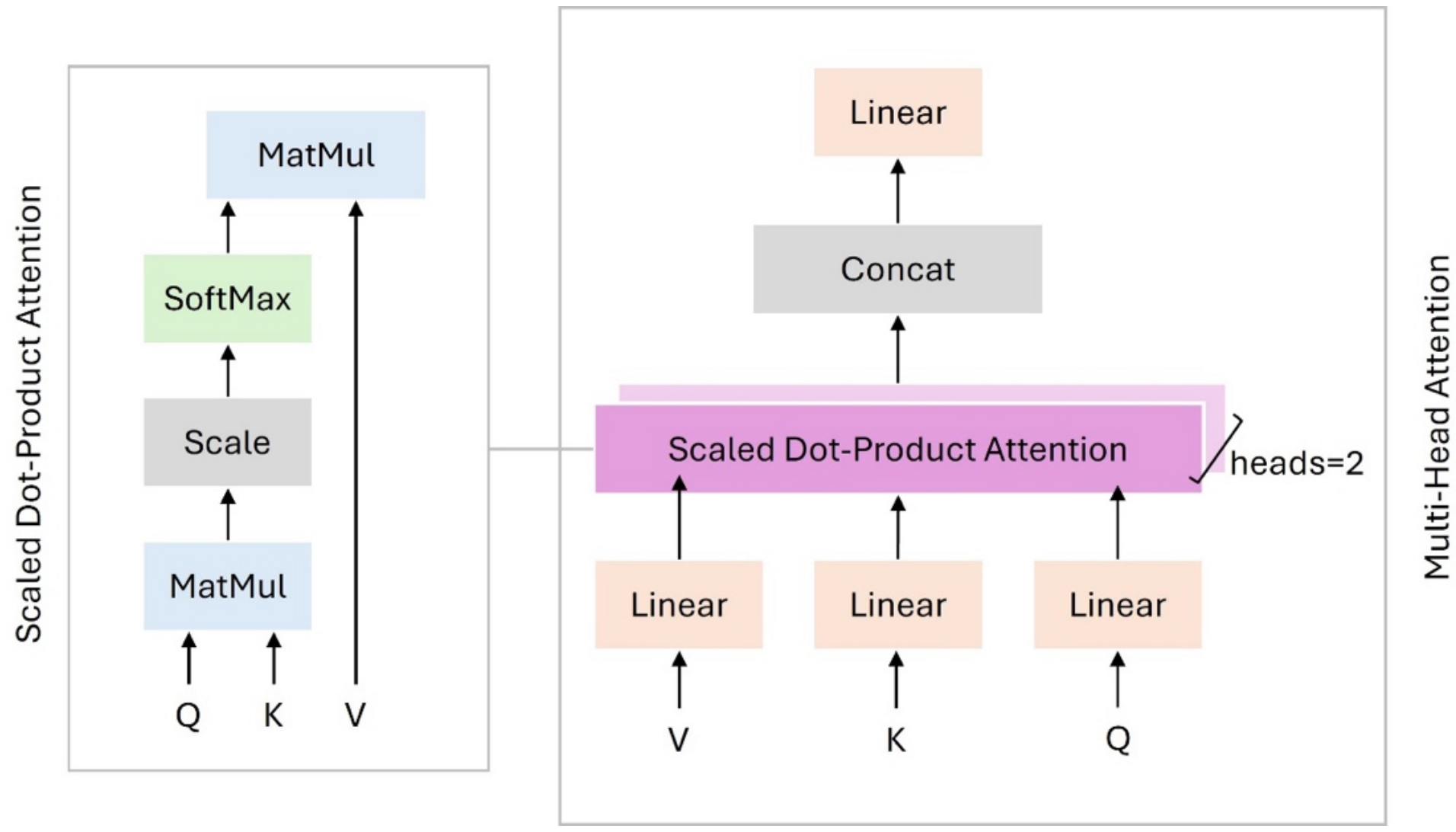

Figure 4. Scaled Dot-Product Attention (left) and Multi-Head Self Attention (right) mechanisms used in the Transformer encoder.

Given query, key and value matrices:

$$Q = XW_Q, \quad K = XW_Q, \quad V = XW_V \tag{20}$$

Each attention head computes:

$$Attention(Q, K, V) = softmax(\frac{QK^\top}{\sqrt{d_k}})V \tag{21}$$

where $d_k = \frac{d_{model}}{n_{head}} = 32$.

Outputs from the $n_{head}$ parallel heads are concatenated and linearly projected to restore dimensionality $d_{model}$.

The attention mechanism learns dynamic, context-dependent weighting of temporal features, allowing the model to integrate distributed neural activations that co-occur across different time scales of motor planning.

Each token (time step) is subsequently processed independently by a non-linear feed-forward network:

$$FFN(x) = \max(0, xW_1 + b_1)\, W_2 + b_2 \tag{22}$$

with $W_1 \in \mathbb{R}^{d_{model} \times d_{ff}}$ and $W_2 \in \mathbb{R}^{d_{ff} \times d_{model}}$.

This layer enhances non-linear mixing and increases representational power while maintaining temporal alignment, since it operates independently across time.

Because self-attention is order-invariant, a rotary positional embedding (RoPE) or sinusoidal encoding $P_t$ is added to each time step to encode relative temporal order:

$$\widetilde{X_t} = X_t + P_t, \quad t = 1, \ldots, T \tag{23}$$

In practice, RoPE better preserves relative phase information in oscillatory EEG signals, aligning the model with neural representations of temporal context.

The output of the $N_L$ stacked encoder layer is denoted

$$H_{global} = TransformerEncoder(H_{local}) \tag{24}$$

where $H_{local}$ is the feature sequence produced by the preceding TCN encoder.

In the Predictive-Coding TCN-Transformer, $H_{global}$ serves as a top-down prior. It predicts the local state of the lower TCN layer through a learned feedback projection:

$$\widehat{H}_{local} = f_{pred}(H_{global}) \tag{25}$$

and the prediction error $E = \widehat{H}_{local} - H_{local}$ is used to iteratively refine both representations (see Predictive Coding subsection).

After encoding, the resulting sequence $H_{global} \in \mathbb{R}^{B \times T \times d_{model}}$ maintains the same temporal length as its input. Each vector encapsulates information from all time steps, modulated by attention weights.

This sequence is then fed into the feedback predictor (top-down refinement), and the $TCN_2$ decoder head for final temporal aggregation.

### 3.4.4 Predictive Coding

In biological neural systems, perception and action are governed by a predictive coding principle: higher cortical regions generate expectations of incoming sensory states, while lower regions compute and propagate prediction errors reflecting mismatches between predicted and observed inputs. This reciprocal flow of information forms a continuous inference process that minimizes free energy or surprise, maintaining adaptive alignment between internal models and the external world.

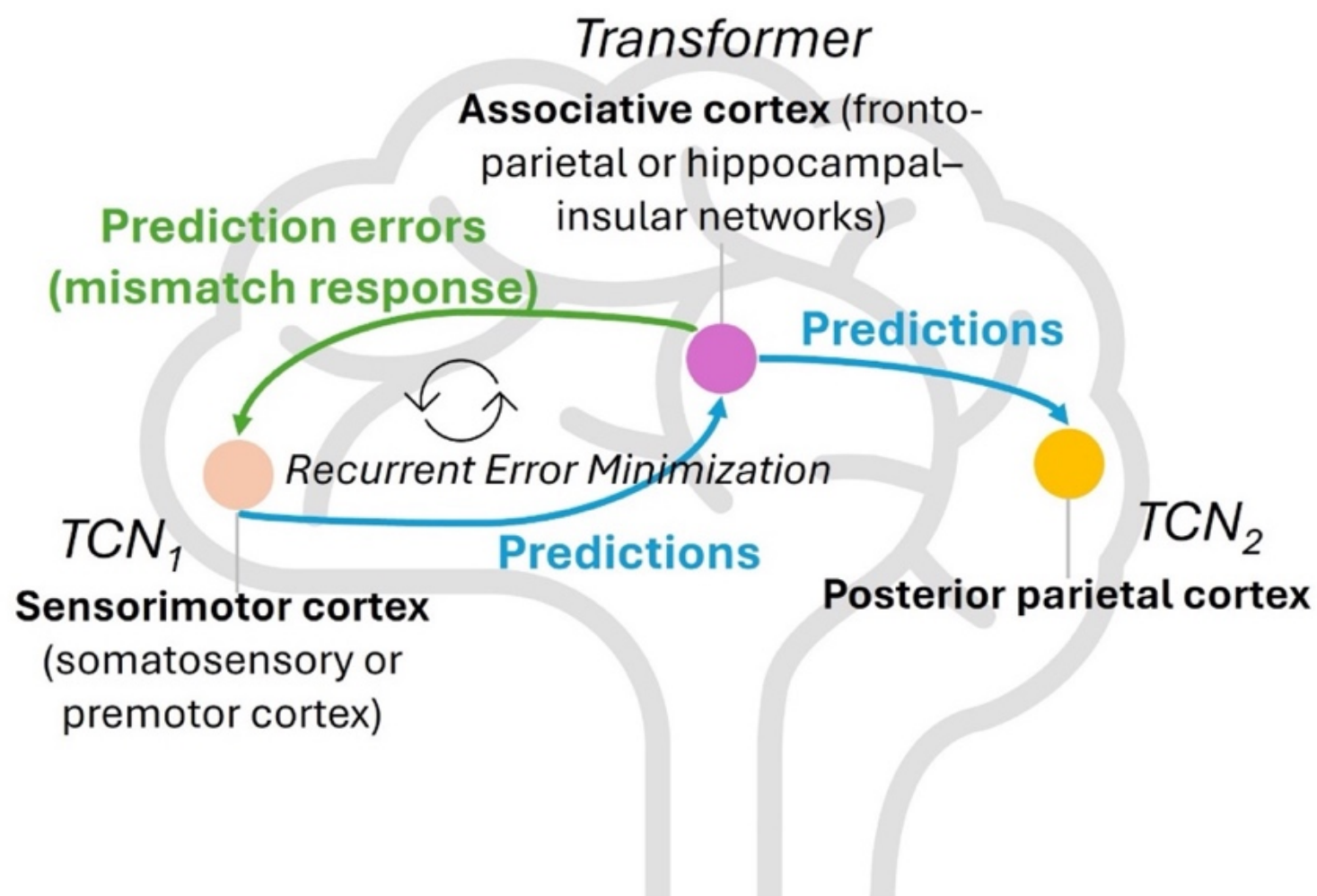

Figure 5. Neurocomputational analogy of the proposed predictive-coding architecture.

Inspired by this framework, the proposed Predictive-Coding TCN-Transformer implements a computational analogue of this bidirectional interaction between hierarchical levels of processing. As shown in Figure 5, the Transformer encoder represents the *high-level generative model*, integrating global temporal and contextual patterns, whereas the first TCN encoder ($TCN_1$) functions as the *low-level sensory model*, encoding fine-grained local dynamics of EEG signals. Predictive feedback from the Transformer to the TCN drives iterative refinement toward a self-consistent representation of motor-planning dynamics.

Let:

- $H_{local}^{(0)} \in \mathbb{R}^{B\times C_1\times T}$ be the local features extracted by $TCN_1$;
- $H_{global}^{(k)} \in \mathbb{R}^{B\times T\times d_{model}}$ be the global contextual features produced by the Transformer at iteration $k$;
- $H_{pred}^{(k)}$ be the predicted lower-layer state inferred from the global representation.

(1) Top-down prediction

The Transformer output is projected back to the TCN feature space through a learned feedback predictor $f_{pred}$:

$$H_{pred}^{(k)} = f_{pred}(H_{global}^{(k)}) \tag{26}$$

implemented as a 1×1 convolution followed by a non-linear activation (tanh).

This projection produces a synthetic reconstruction of the expected local activity pattern given the current global model.

(2) Error computation

The prediction error at iteration $k$ is computed element-wise:

$$E^{(k)} = H_{pred}^{(k)} - H_{local}^{(0)} \tag{27}$$

This term represents the discrepancy between top-down expectations and bottom-up sensory encoding.

(3) Adaptive gain and gating

Not all feedback corrections are equally reliable. An adaptive gating network g(·) regulates the magnitude of error-driven updates:

$$G^{(k)} = \sigma(g(mean_T(H_{global}^{(k)}))) \tag{28}$$

where $\sigma$ is a sigmoid function mapping to [0,1].

The resulting gain $G^{(k)} \in \mathbb{R}^{B\times C_1\times T}$ weights each channel's correction strength.

(4) Bottom-up update

The local representation is then updated through a gated prediction-error correction:

$$H_{local}^{(k+1)} = H_{local}^{(k)} + G^{(k)} \odot E^{(k)} \tag{29}$$

where ⊙ denotes element-wise multiplication.

This iterative refinement continues for a fixed number of feedback cycles ($K$=1–3 in our experiments).

(5) Re-encoding

The corrected local representation is reprojected and re-encoded through the Transformer:

$$H_{global}^{(k+1)} = TransformerEncoder(Proj(H_{local}^{(k+1)})) \tag{30}$$

closing the predictive loop and enabling convergence toward a stable joint representation.

This bidirectional interaction can be interpreted as an *approximate Bayesian inference* process. At each iteration, the top-down prediction minimizes the variational free energy:

$$\mathcal{F} = \underbrace{\left\| H_{pred}^{(k)} - H_{local}^{(k)} \right\|_2^2}_{prediction\ error} + \underbrace{\lambda KL\left[ q\left(H_{global}^{(k)}\right) \parallel p(H_{global}) \right]}_{model\ complexity\ term} \tag{31}$$

where the first term enforces local consistency and the second regularizes global representation stability.

Although the system is trained via backpropagation rather than explicit variational optimization, the recurrent update dynamically drives the hidden states toward a *low-error manifold* in which global and local codes are mutually predictive.

After $K$ feedback iterations, the final local representation $H_{local}^{(K)}$ encapsulates both sensory fidelity and contextual prediction. This representation is then processed by the second temporal convolutional network ($TCN_2$) to capture fine temporal dependencies and generate the final output.

## 3.5 Evaluation method, baselines, and performance metrics

### 3.5.1 Data preparation and preprocessing

EEG recordings were first preprocessed and segmented into fixed-length temporal windows. A sliding window approach was adopted with an exhaustive search method. The length and stride of each window started at 0.25 seconds (s) and incrementally increased by 0.125s at each step until a maximum of 2s. As a result, the total of windows $W$ extracted from the time series is given by

$$W = \left\lfloor \frac{N-L}{S} \right\rfloor + 1 \tag{32}$$

where $N$ is the total number of data points, $L$ is the window length, $S$ is the stride. $\lfloor * \rfloor$ denotes the floor function that returns the greatest integer less than or equal to the given value.

The last ($W$-th) window, $\omega_W$, can be described as:

$$\omega_W = \{x_{t_W}, x_{t_{W+1}}, x_{t_W}, \dots, x_{t_{W+L-1}}\} \tag{33}$$

Given the prediction task is will the subject decide to cross within the next $\Delta L$ seconds, this last segment, which reaches the end of the sequence, was assigned a label as positive. To mitigate inter-subject variability and ensure consistency across sessions, all trials were normalized channel-wise to zero mean and unit variance.

To ensure statistical robustness, the dataset was partitioned into training and testing subsets under a k-fold cross-validation scheme (with $k$=5), where each fold served as a held-out test set once while the remaining folds were used for model training and hyperparameter optimization. Given the naturally unbalanced distribution between positive and negative segments, the training data were augmented through Adaptive Synthetic Sampling Approach (ADASYN) (He et al., 2008) of the minority class to achieve approximate balance in class representation. This step ensures stable gradient updates and prevents bias toward majority patterns.

### 3.5.2 Baseline methods

To ensure a rigorous and interpretable evaluation, we benchmarked our proposed architecture against fifteen baseline classifiers drawn from established literature on EEG decoding, temporal modeling, and sequence representation learning, including traditional distance-based methods, convolutional baselines, and hybrid deep temporal networks. These baselines serve as comparative references for performance benchmarking and form the basis for a systematic ablation analysis, allowing us to isolate and evaluate the contribution of each architectural component, such as causal convolution, attention integration, and predictive feedback, within our proposed predictive-coding framework.

These baselines can be grouped as follows:

1. Non-deep baseline
    - DTW-KNN: A classical dynamic time warping kernel (Sakoe & Chiba, 1978) combined with k-nearest neighbors (k=5) served as a non-parametric reference model. This baseline measures sequence similarity directly in the temporal domain, providing a transparent distance-based comparison.
    - SVM-RBF: A Support Vector Machine with a Radial Basis Function kernel was employed to evaluate the performance of a kernelized boundary learner that does not explicitly model temporal dependencies.
    - Decision Tree: A CART-based decision tree classifier was included to assess performance under interpretable, rule-based partitioning. This model serves as a lower-complexity reference for nonlinear decision surfaces derived from instantaneous features.
    - XGBoost: An Extreme Gradient Boosting ensemble, which iteratively builds additive regression trees optimized via second-order gradient updates, capturing nonlinear dependencies through feature interactions without sequential modeling.
2. Standard deep temporal networks
    - BiLSTM, Attn-LSTM: Sequential recurrent baselines modeling temporal dependencies with varying directionality and attention mechanisms.
    - GRU: A lightweight recurrent alternative to LSTM, serving as a benchmark for gated temporal modeling.
3. Temporal Convolutional Networks (TCN)

- Base-TCN (Ingolfsson et al., 2020): A canonical dilated causal convolution model.
- Gated-TCN, TCN-GAP, DS-TCN-SE (Zhang et al., 2024), Dual-TCN: Architectural variants exploring gating mechanisms, global average pooling, depthwise separable convolution with squeeze–excitation, and dual-stack structures, respectively.

4. Transformer hybrid variants
   - TCN-Transformer (Su et al., 2024), Transformer-TCN (Pesen & Altuncu, 2025), TCN-Transformer-TCN: Hierarchical hybrids integrating convolutional encoders with Transformer-based global attention for multi-scale temporal reasoning.
   - PredictiveCoding-TCN-Transformer (ours): A biologically inspired hybrid architecture integrating top-down predictive feedback and iterative error minimization between Transformer-based contextual inference and TCN-based local encoding.

All baseline implementations were reproduced from publicly available code or re-implemented following their original descriptions. All models were trained under identical preprocessing, cross-validation, and evaluation conditions to ensure fair comparison.

#### 3.5.3 Training objective

Each neural classifier was optimized to minimize the binary cross-entropy loss between the predicted and true class labels:

$$\mathcal{L} = -\frac{1}{N}\sum_{i=1}^{N}[y_i \log(\hat{y}_i) + (1 - y_i)\log(1 - \hat{y}_i)] \tag{34}$$

where $y_i \in \{0,1\}$ denotes the ground-truth label and $\hat{y}_i \in [0,1]$ represents the predicted probability of the active (intent) class.

Models were trained with AdamW (learning rate $1 \times 10^{-3}$, weight decay $1 \times 10^{-4}$). Mini-batches of size 128 were optimized for up to 30 epochs, with gradient-norm clipping at 1.0 each step to stabilize updates. Early stopping monitored validation loss with a patience of 5 epochs and a minimum improvement threshold of $1 \times 10^{-4}$. Dropout regularization and batch normalization were employed within all deep models to enhance generalization. For each classifier, the model weights from the epoch yielding the best validation performance were selected for testing.

#### 3.5.4 Performance metric

Given the imbalanced nature of the EEG dataset, where the distribution between intention and non-intention segments is uneven, accuracy alone may not reliably reflect true discriminative performance. Therefore, we adopted the Area Under the Receiver Operating Characteristic Curve (AUROC) as the primary evaluation metric.

The AUROC measures the model's ability to distinguish between positive and negative classes across all classification thresholds:

$$AUC = \int_0^1 TPR(FPR^{-1}(x))dx \tag{35}$$

where $TPR$ (true positive rate) and $FPR$ (false positive rate) quantify sensitivity and fall-out, respectively. An AUC of 0.5 indicates random guessing, while 1.0 denotes perfect separability. This metric is particularly suitable for imbalanced EEG classification tasks, as it remains invariant to label proportions and focuses on ranking quality rather than absolute classification frequency.

For completeness, we additionally report mean and standard deviation of the AUC values across the five cross-validation folds, providing a measure of both performance and robustness for each baseline model.

#### 3.5.5 Experimental setup and data size

All experiments were conducted using the processed EEG dataset collected from 12 participants, each completing five experimental trials. Due to the relatively small dataset, signals from all participants were aggregated into a single training pool rather than adopting a subject-specific partitioning scheme. This design choice ensures sufficient data coverage for model convergence and allows the learning algorithm to generalize across heterogeneous neural signatures. The heterogeneity inherent in the dataset from inter-individual variability in brain dynamics, electrode impedance, and task adaptation is advantageous for evaluating model robustness and transferability in realistic, multi-user BCI scenarios.

Each continuous EEG recording was segmented using a sliding-window approach, where the window length and stride were determined empirically to balance temporal resolution and sample independence. Shorter strides increased the number of overlapping segments, providing more training instances while preserving fine-grained temporal dynamics. The final dataset size therefore varied with the chosen windowing configuration. For example, using a sliding window of 9 steps with a stride of 3 steps produced 838 segments, whereas a window of 11 steps with a stride of 7 steps yielded 356 segments. Despite these differences, all configurations maintained consistent coverage of both positive and negative intention phases.

## 4 Results and discussion

### 4.1 Feature selection

Drawing upon the Perception–Action Model (Kowalski-Trakofler & Barrett, 2003; Windridge et al., 2013), we hypothesized that the pedestrian road-crossing decision process unfolds through four latent cognitive microstates (Tian et al., 2022):

(1) *Perceiving the environment*—selective attention to salient visual cues such as oncoming vehicles;

(2) *Assessing risk*—evaluating vehicle speed and distance to estimate collision likelihood;

(3) *Determining the available time to cross*—integrating temporal and spatial information to assess safety margins;

(4) *Initiating movement*—executing the final decision to step into the roadway.

These cognitive microstates reflect a hierarchical progression from perceptual encoding to motor preparation, engaging high-level cognitive functions such as selective attention, which orients focus toward relevant stimuli (Ptak, 2012), and working memory, which facilitates rule-based evaluation and response selection (Bunge, 2005).

To empirically identify the temporal boundaries of these hypothesized states, we employed a Hidden Markov Model (HMM) (Rabiner, 1989) to perform probabilistic inference over the multivariate EEG time series. In this model, the pedestrian's decision-making process is conceptualized as a Markov process with four hidden states, representing the latent cognitive microstates that generate the observed EEG spectral dynamics.

The HMM was trained on the EEG power spectral data, leveraging the sequential dependencies among latent cognitive states. Parameter estimation was achieved through the Expectation–Maximization (EM) procedure, specifically the Baum–Welch algorithm (Fine et al., 1998), which iteratively maximized the log-likelihood function:

$$\mathcal{L}(\theta) = \sum_{t=1}^{T} log \sum_{s_t} P(x_t|s_t,\theta)P(s_t|s_{t-1},\theta) \tag{36}$$

where $x_t$ denotes the observed spectral feature vector at time $t$, $s_t$ the hidden state, and $\theta$ the model parameters (transition and emission probabilities). Iterations continued until convergence, defined as a change in log-likelihood below a threshold ($\Delta\mathcal{L} < 10^{-4}$), ensuring parameter stability.

After model optimization, the Viterbi algorithm was used to infer the most probable sequence of hidden states:

$$\hat{s}_{1:T} = arg\max_{s_{1:T}} P(s_{1:T}|x_{1:T},\theta) \tag{37}$$

thereby reconstructing the temporal sequence of cognitive microstates underlying each trial. Figure 6(a) illustrates the resultant temporal structure, showing the transitions and duration of each inferred microstate across the decision-making timeline.

Prior to HMM training, the 70-dimensional EEG data were standardized to zero mean and unit variance to eliminate scaling effects. The data were then reduced using Principal Component Analysis (PCA) (Yu et al., 2014) to minimize noise and redundancy while preserving key variance components. The top five principal components collectively explained 94% of the total variance, as shown in Figure 6(b).

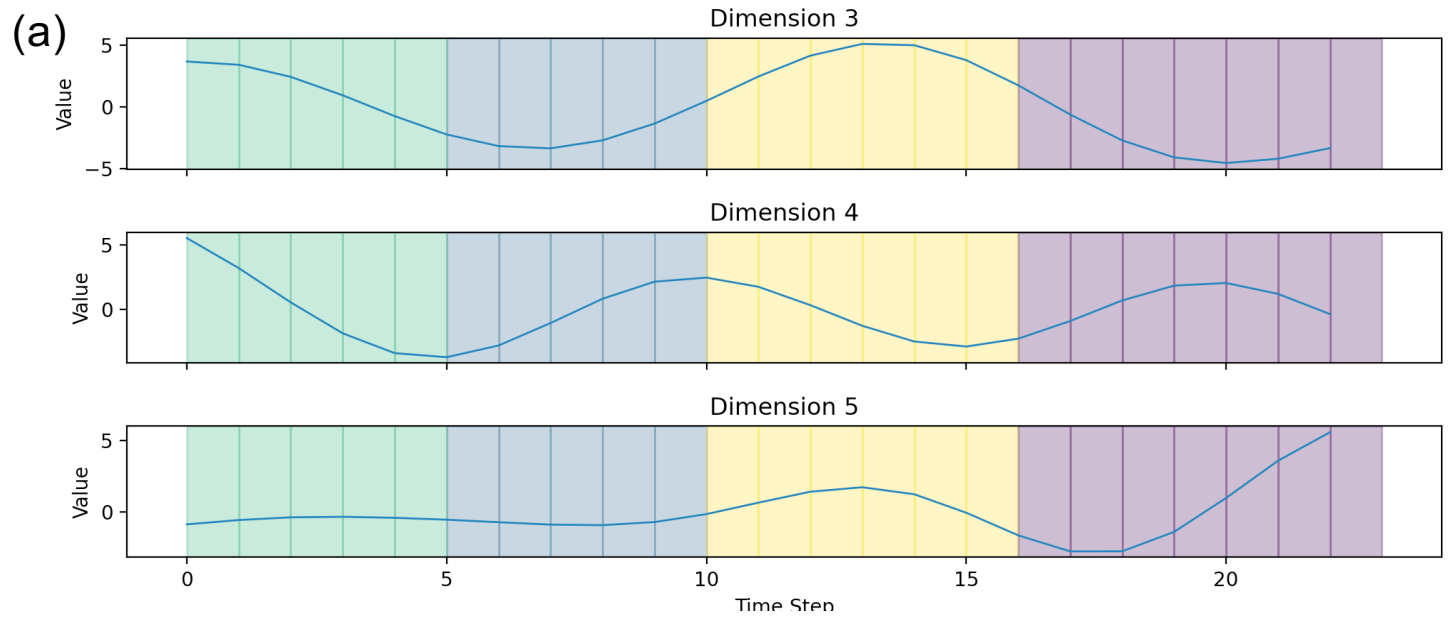

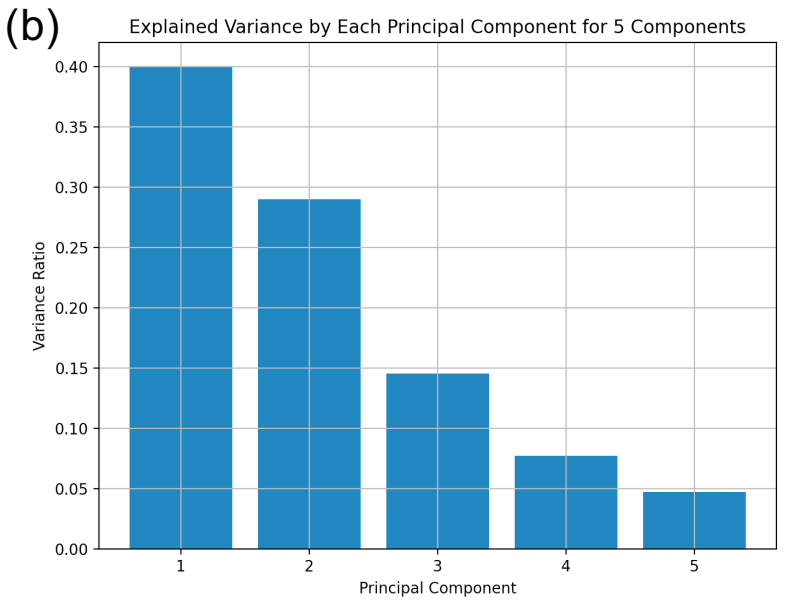

Figure 6. (a) Cognitive microstates decomposed by the Hidden Markov Model (HMM) from the principal component representations of EEG data, illustrated for a representative trial. (b) Variance explained by the top five principal components derived from principal component analysis (PCA).

The Gaussian HMM was subsequently applied to the reduced feature space on a trial-by-trial basis. Upon convergence, the posterior state probabilities were used to assign each time point to one of the four inferred microstates. Each trial was then segmented into four temporal windows corresponding to the predicted microstate transitions. To mitigate the influence of statistical outliers, values falling outside the interquartile range (IQR), defined as $[Q_1 - 1.5 \times IQR, Q_3 + 1.5 \times IQR]$, were excluded from further analysis.

Since the PSD distributions for each microstate did not satisfy normality assumptions (Shapiro–Wilk $p > 0.05$), we adopted Friedman's Analysis of Variance (ANOVA) (Friedman, 1937), a non-parametric alternative to repeated-measures ANOVA, to assess differences in EEG power across the four microstates. Following significant omnibus results, Conover's post-hoc tests were conducted to determine pairwise contrasts between states. All statistical tests were two-tailed with a 5% significance threshold. $p$ -values were used to assess significance, while t-scores indicated the direction and magnitude of mean differences. Cohen's $d$ was computed to quantify effect sizes, with $d \geq 0.4$ considered the minimal threshold for reliable psychological effects (Brysbaert, 2019).

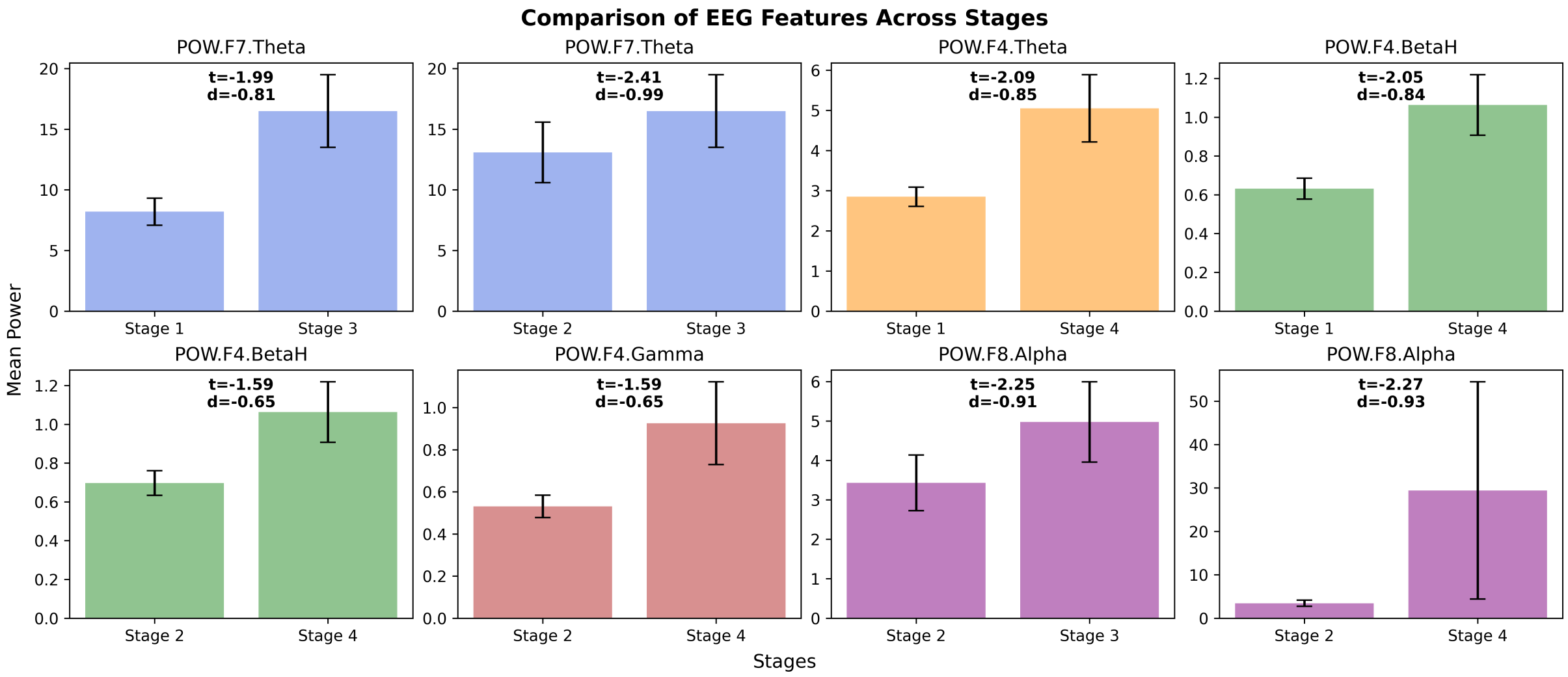

Figure 7. Post-hoc pairwise comparisons showing significant differences across microstates, reported with $t$-values and Cohen's $d$ effect sizes. Error bars represent standard deviations. Colors denote different EEG feature dimensions.

Figure 7 highlights the spectral–spatial features showing significant differentiation across microstates, primarily localized in frontal regions, indicating strong engagement of executive control areas.

- In the perceptual stage, increased theta-band activity at the left frontal electrode (F7) was observed, consistent with its established role in initial perceptual encoding and environmental monitoring (Karakaş, 2020).

- Transitioning to selective attention, alpha-band event-related desynchronization (ERD) emerged in the right frontal region (F8), reflecting inhibitory gating that suppresses irrelevant sensory input to facilitate focused attention (Foxe & Snyder, 2011).
- During the motor planning phase, pronounced high-beta oscillations were identified in the right frontal cortex (F4), corroborating previous evidence linking beta-band activity with motor preparation and response initiation (Wagner et al., 2017).
- Finally, during decision execution, elevated gamma-band activity was detected, representing high-order cognitive control processes underlying decision resolution and motor initiation (Jensen et al., 2007; Polanía et al., 2012).

Together, these results delineate a hierarchical perceptual–cognitive–motor sequence, where EEG dynamics evolve from broad environmental surveillance to focused attentional evaluation and finally to motor commitment. The identified spectral–spatial markers thus provide a neurophysiologically grounded basis for feature extraction in the subsequent predictive modeling.

### 4.2 EEG analysis

To corroborate the neurophysiological validity of the recorded EEG data and ensure that the extracted features accurately represent cognitive processes underlying perceptual–motor decision-making, we conducted a detailed multiscale analysis combining event-related potential inspection, spectral–temporal decomposition, and functional connectivity mapping. The findings collectively indicate that the recorded EEG signals are of sufficient spatial and temporal resolution to capture biologically plausible neural dynamics involved in perception, attention, and motor planning during pedestrian crossing decisions.

#### 4.2.1 Event-related potentials and spatial activation

Figure 8 illustrates the butterfly plots of raw EEG traces, overlaid with scalp topographies of event-related brain activity at successive time intervals. Topographies were extracted at four equidistant timepoints that collectively represent the progression from perception to action. The representative examples in Figures 8(a) and 8(b) correspond to trials with total response times of 2.145 s and 2.845 s, respectively.

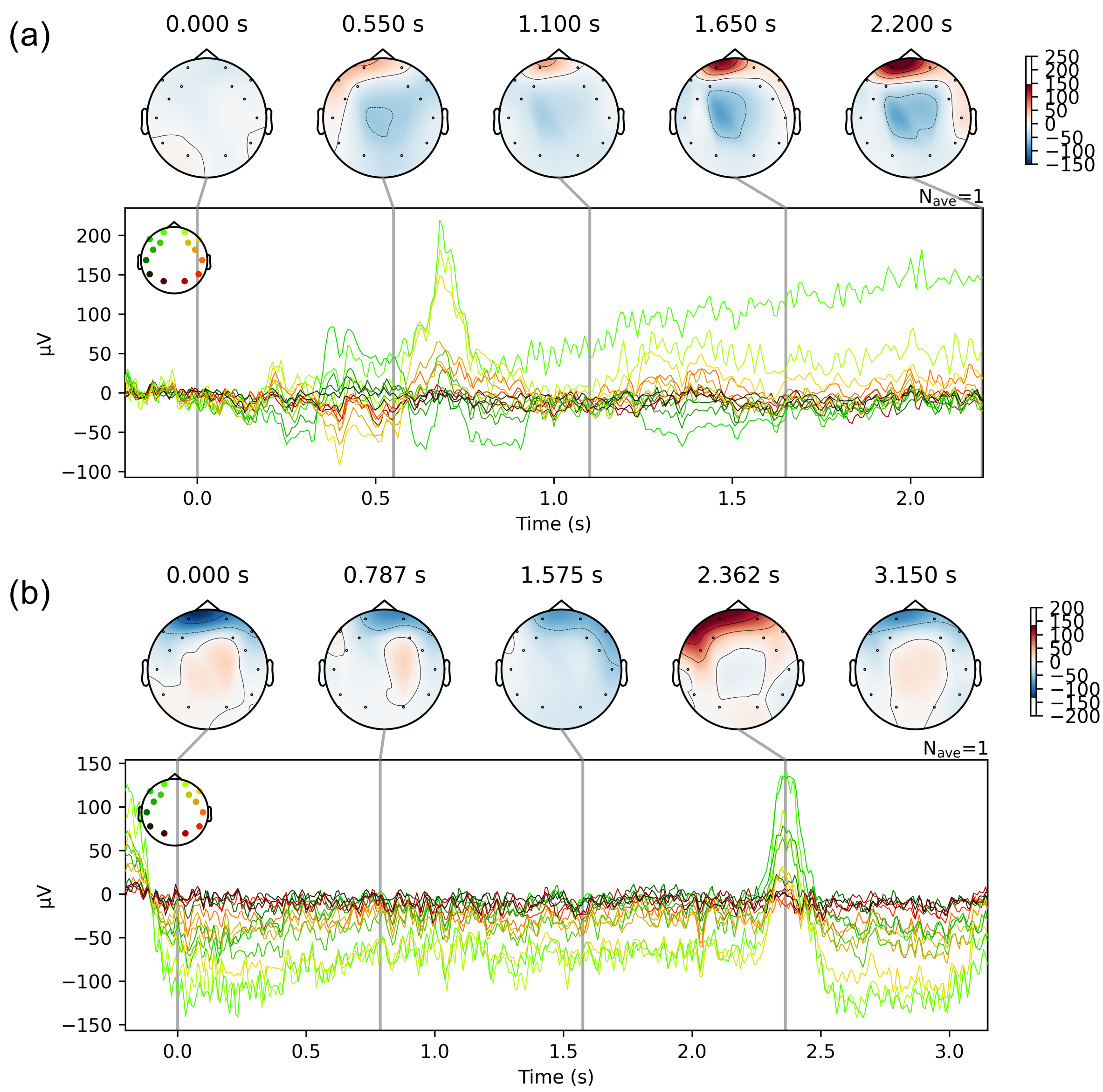

Figure 8. Joint plots combining butterfly traces with scalp topographies for (a) the *Low Traffic Volume* scenario and (b) the *Signalized Intersection* scenario.

Across participants, a consistent spatiotemporal pattern emerged: approximately 0.5 s prior to the behavioral response, pronounced positive deflections (increased amplitude) were observed over left frontal electrodes, while preceding intervals showed low or negative activation in the same region. This temporal buildup of frontal activation likely reflects motor preparedness and response readiness, consistent with the role of the left prefrontal cortex in voluntary action initiation (Sulpizio et al., 2017). The sequential topographies thus reveal a transition from diffuse, low-level activation during early perception to focal, high-amplitude frontal engagement immediately preceding the motor response.

#### 4.2.2 Spectral characteristics and frequency–spatial distribution

The PSD distribution across all channels (Figure 9a) revealed the canonical 1/f spectral profile, in which low frequencies exhibit higher power while high-frequency components attenuate gradually. This is consistent with the known scale-free dynamics of human EEG (He, 2014), confirming the signal's physiological authenticity.

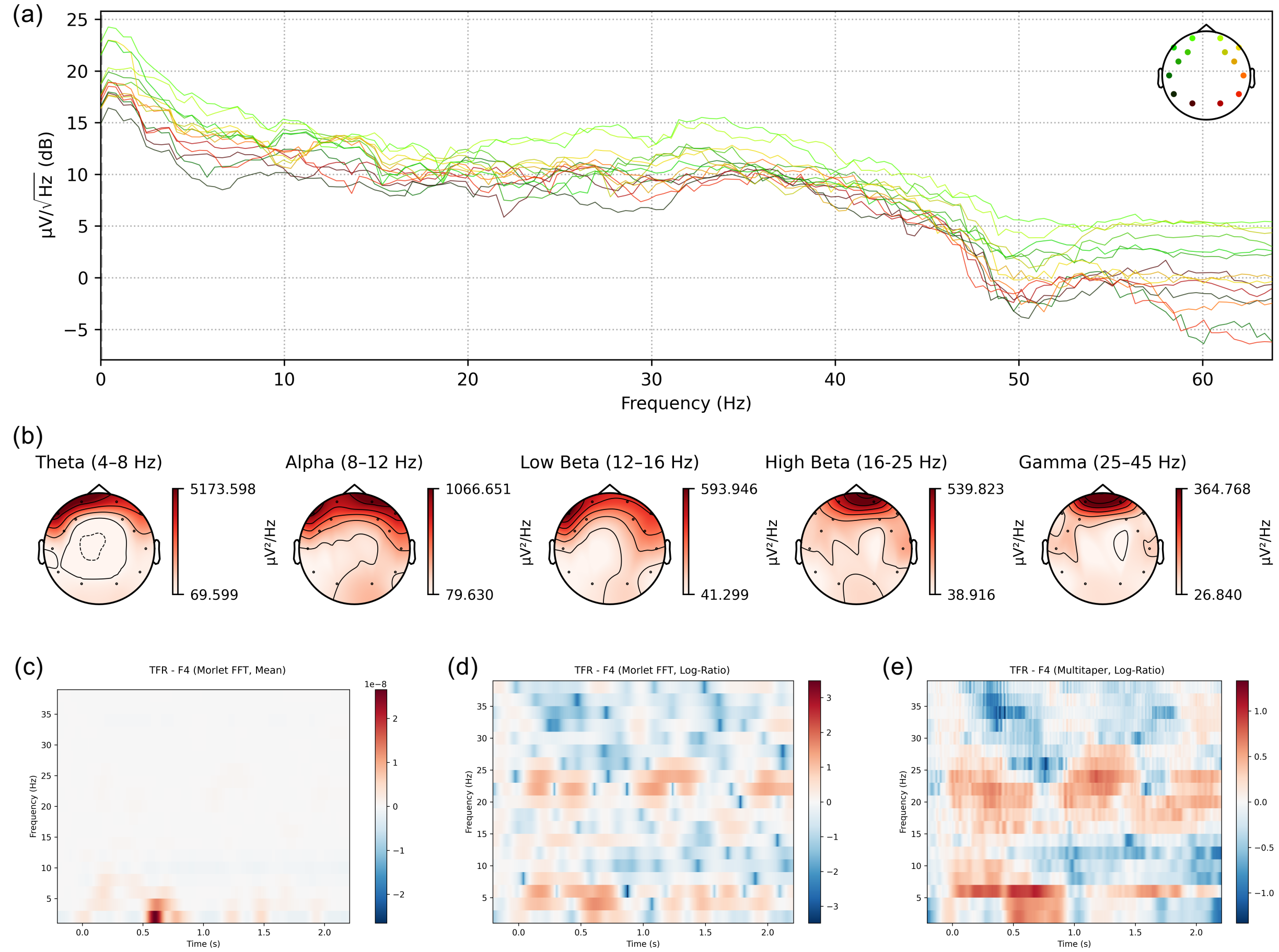

Figure 9. (a) Power Spectral Density across frequencies (0-65 Hz) over 14 EEG channels. (b) Interpolated scalp topography of power in five specific frequency bands. (c-e) Time-Frequency Representations (TFRs) at electrode F4 using Morlet FFT with a mean baseline correction, Morlet FFT with a log ratio baseline correction, and a Multitaper with a log ratio baseline correction method, respectively.

Spatial mapping of spectral power (Figure 9b) demonstrated a clear frontal dominance, with the theta (4–8 Hz) and alpha (8–12 Hz) bands exhibiting the strongest amplitudes relative to higher frequencies. This frontal predominance suggests active top-down cognitive control and attentional engagement, which are typically reflected in low-frequency oscillatory components during perceptual decision-making (Helfrich et al., 2017).

To examine temporal evolution in greater detail, TFRs were computed at electrode F4 using three spectral estimation methods (Figures 10c–e). The Morlet transform achieved superior temporal precision, revealing transient low-frequency power bursts (~0.6 s post-stimulus), whereas the multitaper method produced smoother and more stable frequency contours with improved spectral resolution. Log-transformed normalization preserved higher-frequency activity, enabling clear identification of sustained high-beta (16–25 Hz) oscillations extending from mid-trial through the motor preparation stage. This sustained beta activity suggests the maintenance of motor readiness once an internal decision threshold was reached (Engel & Fries, 2010).

Overall, the spectral evidence indicates a frequency-specific cascade: dominant theta and alpha oscillations during perception and evidence accumulation, followed by high-beta activation associated with motor commitment and execution.

### 4.2.3 Oscillatory dynamics across time and space

Figure 10 provides a temporally and spatially resolved overview of oscillatory effects across the decision-making sequence. During the early perception stage (~0.65 s post-stimulus), the scalp topographies reveal left-lateralized frontal theta dominance, coupled with parietal–temporal suppression. This spatial pattern is characteristic of theta-mediated sensory processing and attentional control (Karakaş, 2020), indicating engagement of frontal networks in filtering and integrating visual input from the environment.

As time progresses toward motor preparation (~2.0 s post-stimulus, corresponding to the average response latency of 2.145 s), activation shifts to high-beta dominance in frontocentral regions, implicating motor-related areas such as the supplementary motor cortex and premotor regions in decision commitment and response readiness (Wagner et al., 2017). Concurrent low-frequency frontal activity persisted during perception, reflecting top-down modulation of posterior sensory processing pathways (Helfrich et al., 2017).

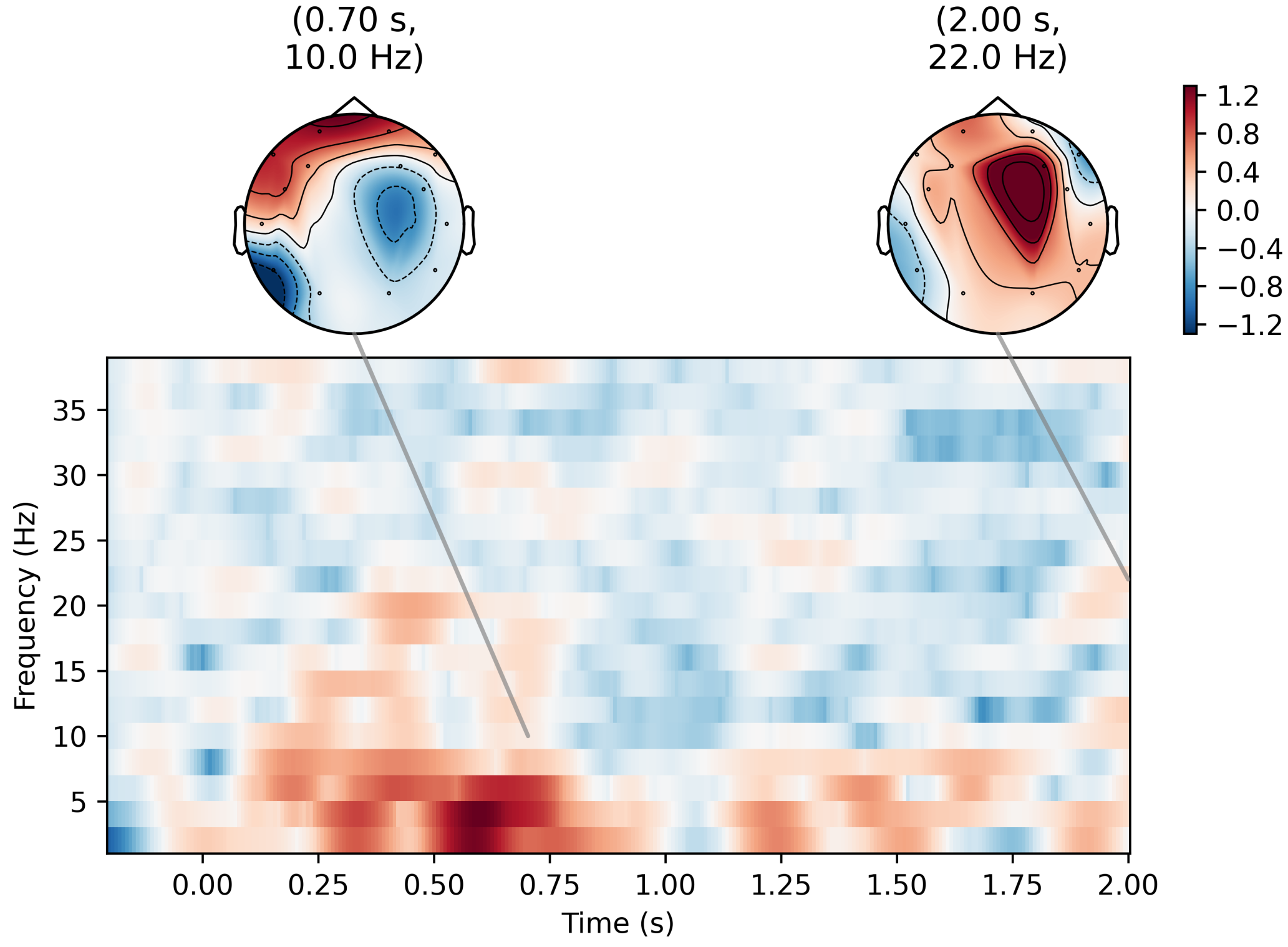

Figure 10. Joint plot showing the aggregated TFR across all EEG channels, along with scalp topographies at two representative times and frequency bands.

### 4.2.4 Functional connectivity and network reorganization

To further examine how cognitive processing unfolds across distributed neural circuits, we computed the wPLI across time and frequency bands to characterize functional connectivity (see Figures 11-12).

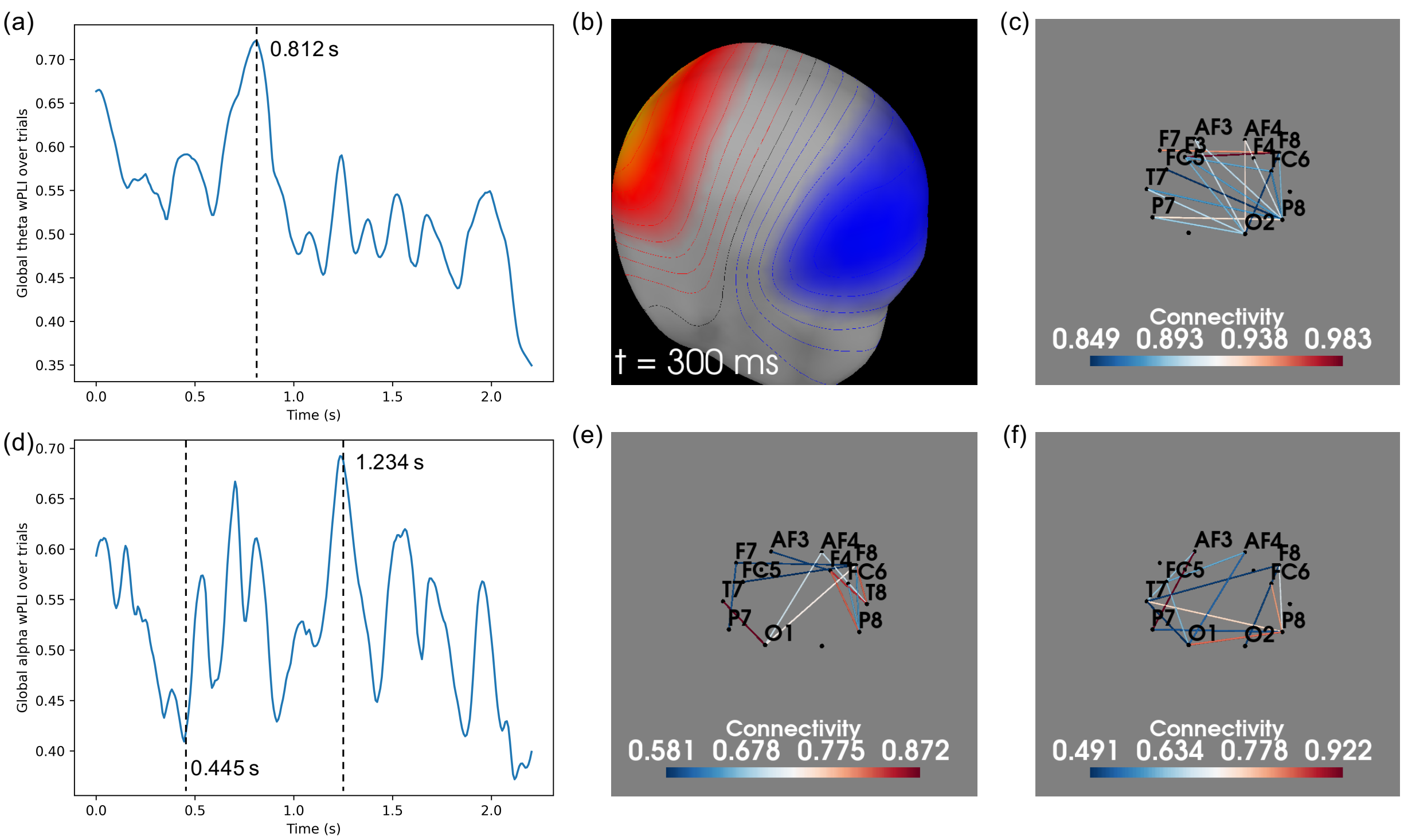

Figure 11. (a),(d) Global wPLI for theta and alpha bands. (b) 3D Field Map at 0.3s post-stimulus. (c),(e),(f) sensor connectivity for theta (0s pre-stimulus), alpha (0.445 post-stimulus), and high beta (0.055 post-stimulus).

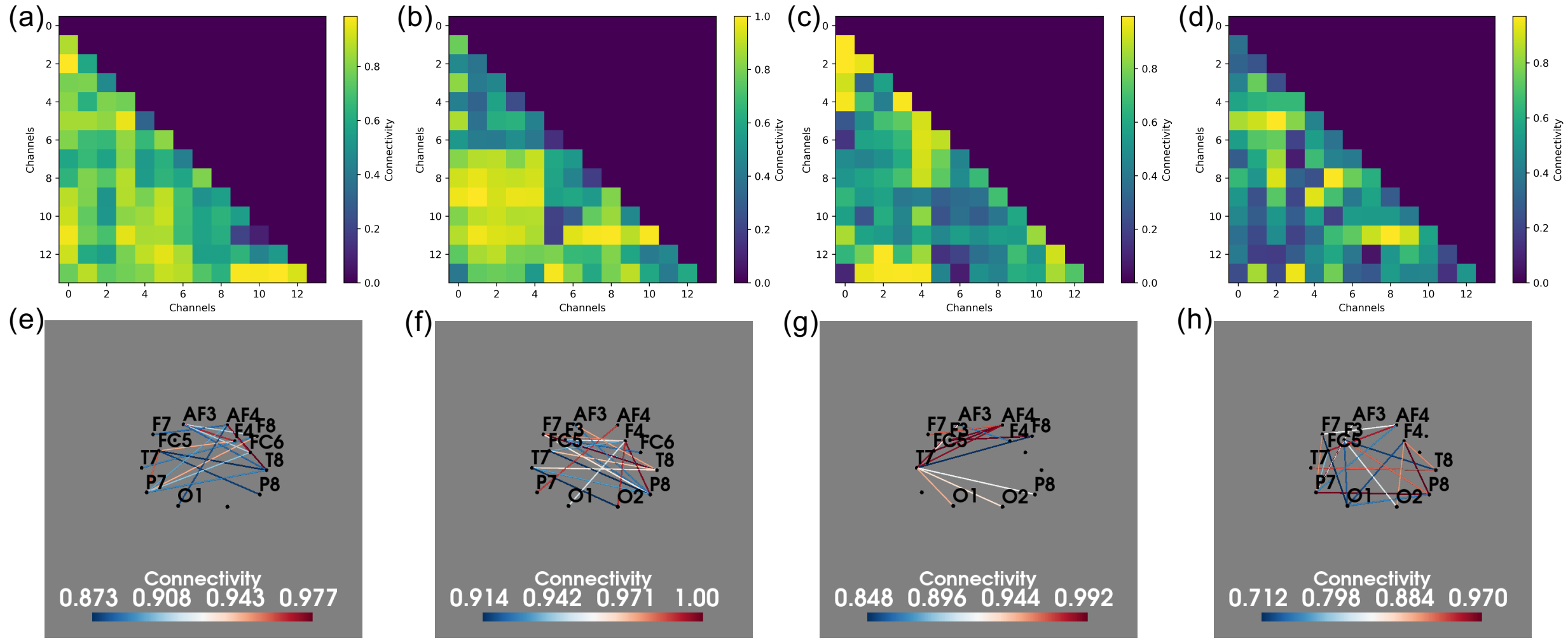

Figure 12. (a)-(d) Sensor connectivity matrices for theta (0.812s post-stimulus), alpha (1.234s post-stimulus), alpha (2s post-stimulus, high-beta (2s post-stimulus.) (e)-(h) Corresponding connectivity visualizations for the above matrices.

At approximately 0.812 s post-stimulus, global theta-band coupling peaked, as shown in Figure 11(a). Connectivity matrices at this timepoint revealed strong cross-hemispheric interactions, initially emerging between the left frontal and right posterior regions (P8, O2). As perception evolved, connectivity reorganized toward a right frontal–left posterior pattern (P7, O1), suggesting dynamic hemispheric reconfiguration of attention-related networks.

This reorganization aligns with prior findings on hemispheric asymmetry, wherein the right hemisphere is preferentially engaged in spatial reasoning and visuospatial integration (Sun & Walsh, 2006). The initial left-frontal dominance likely corresponds to proactive inhibitory control mechanisms activated before stimulus onset and released at stimulus appearance (Sulpizio et al., 2017). The subsequent shift to right-frontal dominance indicates enhanced bottom-up, stimulus-driven processing under perceptual uncertainty (Tsumura et al., 2021).

In the alpha band, connectivity analyses across three key timepoints—0.445 s (lowest wPLI), 1.234 s (global wPLI peak), and 2.0 s (motor stage)—revealed a pattern of early alpha suppression followed by later alpha enhancement. The early suppression signifies active sensory encoding, while the subsequent rebound reflects inhibitory gating that suppresses irrelevant sensory input as the participant transitions from perception to action (Filippi et al., 2020).

For high-beta oscillations, connectivity patterns also demonstrated a distinct transition between perception and action. At early stages (0.055 s post-stimulus), motor-related frontal regions (F4) were absent from the network. By contrast, at 2.0 s post-stimulus, F4 exhibited strong phase coupling with posterior parietal electrodes (P8), consistent with a sensorimotor integration network supporting goal-directed actions (Martínez-Vázquez & Gail, 2018).

#### 4.2.5 Neurophysiological interpretation

Together, these results provide convergent evidence that the collected EEG data capture a plausible sequence of neural events spanning perception, attention, and motor preparation.

1. Theta-band synchronization reflects early attentional engagement and sensory integration.
2. Alpha-band suppression and rebound correspond to alternating cycles of perceptual encoding and inhibitory gating.
3. High-beta activation marks the neural correlate of decision commitment and motor readiness.
4. Dynamic cross-hemispheric connectivity indicates flexible reallocation of neural resources between proactive (top-down) and reactive (bottom-up) control mechanisms.

The observed temporal and spatial patterns validated through both oscillatory and connectivity analyses demonstrate that the dataset effectively captures the neurophysiological underpinnings of real-world decision-making. These findings strengthen the argument that the EEG signals used for subsequent predictive modeling are not only statistically sound but also biologically interpretable, reflecting genuine neural computations associated with perceptual evaluation and action initiation.

### 4.3 Predictive model performance and comparative evaluation

Guided by the neurophysiological analyses presented in §4.2, we selected preparatory neural activity mediated by high-beta (16–25 Hz) oscillations at the right fronto-central site (F4) as the feature dimension for intent prediction. This selection aligns with the observed sustained high-beta power that precedes motor commitment and its functional connectivity with posterior parietal sites during movement preparation. Unless otherwise stated, all results reported in this section are obtained using a sliding window of nine samples and a stride of three samples, which define the temporal resolution of the training sequences.

As shown in Table 3, among all models, the Predictive-Coding TCN–Transformer–TCN architecture achieved the highest AUROC (0.727 ± 0.074), outperforming both its base TCN and hybrid Transformer variants. Relative to the structurally closest baseline (TCN–Transformer–TCN without feedback), the predictive-coding variant also yielded a gain, demonstrating the value of top-down prediction and precision-weighted error updates in improving temporal inference and stability.

Table 3. Prediction Results with F4·High-β (window = 9, stride = 3)

| **Model** | **AUROC (± SD)** |
|---|---|
| **Predictive Coding TCN-Transformer-TCN (ours)** | **0.727 (± 0.074)** |
| TCN-Transformer | 0.692 (± 0.108) |
| Gated-TCN | 0.688 (± 0.075) |
| DS-TCN-SE | 0.695 (± 0.147) |
| TCN-GAP | 0.685 (± 0.193) |
| Dual-TCN | 0.697 (± 0.107) |
| TCN | 0.668 (± 0.210) |
| Transformer-TCN | 0.612 (± 0.210) |
| TCN-Transformer-TCN (no predictive loop) | 0.662 (± 0.133) |
| GRU-Seq2Seq | 0.594 (± 0.278) |
| DTW-KNN | 0.587 (± 0.193) |
| Decision Tree | 0.573 (± 0.112) |
| SVM-RBF | 0.603 (± 0.277) |
| BiLSTM | 0.531 (± 0.267) |
| Attn-LSTM | 0.487 (± 0.248) |
| XGBoost | 0.561 (± 0.235) |

The superior performance of TCN-based models reflects their ability to capture short transient dynamics and local sequential dependencies within EEG signals. In contrast, RNNs and attention-only architectures appear less sensitive to rapid cognitive transitions that characterize early intention formation.

To validate that the observed performance improvements stem from meaningful neural features rather than model complexity or random chance, we deliberately chose a non-diagnostic channel/band (AF3·Theta) and repeated the experiment under the same (9, 3) configuration. As summarized in Table 4, model performance dropped substantially across all architectures.

Table 4. Prediction results from random feature choice (AF3·Theta; window = 9, stride = 3)

| **Model** | **AUROC (± SD)** |
|---|---|
| Decision Tree | **0.456 (± 0.079)** |
| Transformer-TCN | 0.428 (± 0.127) |
| TCN-Transformer | 0.421 (± 0.049) |
| TCN-Transformer-TCN | 0.408 (± 0.047) |
| Gated-TCN / Dual-TCN | 0.378 (± 0.109 / 0.121) |
| Predictive Coding TCN-Transformer-TCN | 0.380 (± 0.074) |
| GRU-Seq2Seq | 0.325 (± 0.147) |
| BiLSTM | 0.301 (± 0.111) |
| TCN | 0.307 (± 0.096) |
| SVM-RBF | 0.390 (± 0.211) |
| DTW-KNN | 0.313 (± 0.089) |
| XGBoost | 0.261 (± 0.049) |
| TCN-GAP | 0.250 (± 0.088) |

Here, no sequential model exceeded an AUROC of 0.46, while a simple Decision Tree performed best among an overall weak set of classifiers. This collapse in predictive accuracy confirms that our results do not arise from model ordering or random initialization, but rather from principled feature selection grounded in neurophysiological evidence. The finding also highlights that the F4·High-β feature encapsulates meaningful preparatory neural processes essential for decoding intent.

To further examine the effect of temporal context on model performance, we systematically varied the sliding-window length and stride and evaluated both the DTW-KNN (as a transparent baseline) and the full model set. The results were ranked by the AUROC from the final cross-validation fold and then reported as averaged across all folds.

Table 5. DTW-KNN AUROC vs. windowing

| **(length, stride)** | **AUROC (± SD)** |
|---|---|
| (9, 3) | 0.587 (± 0.193) |
| (11, 7) | **0.760 (± 0.203)** |
| (8, 9) | 0.710 (± 0.074) |
| (5, 9) | 0.679 (± 0.073) |

The configurations reported in Table 5 correspond to the top four performers based on DTW-KNN's last-fold AUROC ranking. As shown, a longer window with a moderate stride (11, 7) yielded the highest DTW-KNN performance, likely because DTW benefits from a richer temporal shape and fewer redundant overlaps, whereas short and overlapping windows (9,3) inflate redundancy and amplify local noise, hurting a non-parametric similarity-based matcher.

When the temporal window was increased to 11 samples with a stride of 7 across all models, a notable reordering of model performance emerged (see Table 6).

Table 6. Prediction Results with F4·High-β (window = 11, stride = 7)

| Model | AUROC (± SD) |
|---|---|
| Attn-LSTM | **0.838 (± 0.112)** |
| GRU-Seq2Seq | 0.840 (± 0.130) |
| TCN-Transformer | 0.821 (± 0.133) |
| TCN-Transformer-TCN | 0.814 (± 0.168) |
| SVM-RBF | 0.812 (± 0.126) |
| XGBoost | 0.791 (± 0.250) |
| **Predictive Coding TCN-Transformer-TCN (ours)** | 0.792 (± 0.198) |
| DTW-KNN | 0.760 (± 0.203) |
| TCN | 0.777 (± 0.210) |
| Gated-TCN | 0.785 (± 0.178) |
| DS-TCN-SE | 0.789 (± 0.229) |
| Transformer-TCN | 0.690 (± 0.231) |
| Decision Tree | **0.663 (± 0.191)** (worst) |

Under this configuration, longer windows expose richer temporal dependencies (multi-hundred-ms context). Models designed to learn temporal dynamics, including RNNs (Attn-LSTM, GRU) and Transformer hybrids capitalize on this added context, hence the overall jump in AUROC and the dominance of sequence learners.

In contrast, Tree ensembles and shallow learners degrade with longer windows: as feature vectors grow more correlated and high-dimensional, they require stronger regularization or manual summarization; without it, they underfit complex temporal structure, as this violates their independence assumptions and limits their ability to represent temporal continuity. (Decision Tree becomes worst).

Our Predictive-Coding TCN–Transformer–TCN model performed modestly under the (11, 7) configuration, where the contribution of the feedback loop was also less evident. Across this setting, all sequence-based architectures, particularly RNNs and attention-integrated models, demonstrated superior performance, confirming that longer temporal spans favor architectures capable of capturing extended dependencies. However, TCN-based models exhibit a distinct advantage in shorter temporal horizons, where rapid temporal dynamics and localized feature interactions dominate. This insight, to our knowledge, has not been explicitly articulated in prior literature. As presented in §2, previous studies have consistently reported TCN and its variants as state-of-the-art performers across various BCI benchmarks when compared with RNN-based methods. The present results clarify that this superiority is particularly pronounced for short, transient neural signals. Although shorter windows may not achieve the same absolute performance or stability as longer ones, they yield near-instantaneous predictions, which are crucial for real-time BCI applications and predictive alerting systems. In this short-window regime, the TCN remains the most effective architecture, and our predictive-coding-enhanced variant provides the best overall performance.

## 5 Conclusion

To help mobile robotic platforms operate safely and efficiently alongside humans in shared built environments, this study proposes humans becoming active beacons for a cooperative sensing paradigm and presents a neurophysiologically grounded framework for decoding human motion intentions from wearable EEG signals using a predictive-coding TCN–Transformer hybrid architecture. The model design embodies a hierarchical interaction between local causal processing and global temporal integration. The TCN progressively expands its temporal receptive field through dilated convolutions, capturing short-horizon causal dependencies that characterize the early buildup of motor-planning dynamics. The Transformer module, operating atop this structure, integrates non-local temporal context through attention-based inference, enabling long-range reasoning across extended decision epochs.

Algorithmically, this hybridization combines the TCN's strength in fine-grained temporal precision with the Transformer's capacity for contextual abstraction. From a neurobiological perspective, the architecture mirrors the interplay between associative cortical regions and sensorimotor circuits. The TCN modules emulate feedforward processing within lower sensory–motor hierarchies, where immediate evidence is accumulated in a time-causal manner, while the Transformer functions analogously to higher-order associative areas (e.g., hippocampal–insular networks) that form generative expectations over sensory evidence. Attention mechanisms distribute contextual weighting in a manner reminiscent of cortical feedback, dynamically regulating lower-level activations.

This synergy allows the model to jointly encode what is happening (semantic representation) and when it occurs (temporal structure), enabling robust inference of latent human intentions from dynamic EEG patterns. The resulting computational loop (prediction, feedback correction, and evidence updating) closely parallels the free-energy minimization process observed in the perception–action cycle of the human brain. The Transformer–TCN interaction thus not only provides algorithmic efficacy but also reflects a biologically plausible mechanism for integrating predictive coding principles into machine learning architectures.

**Limitations and future directions**

Although this study provides proof-of-concept for the proposed framework for predicting motor-planning intentions from EEG data, several limitations warrant further consideration. First, the experiment employed a visual mental paradigm, in which participants made crossing decisions based on observed traffic scenarios on screen and pressed a key to simulate executing physical movements. This design was chosen to ensure experimental control and avoid physical collision risks, while still capturing authentic perceptual–cognitive–motor processing. Importantly, neurophysiological evidence supports the functional equivalence hypothesis that observing, imagining, or executing actions engages overlapping neural substrates. Prior work shows that action observation, motor imagery, and even linguistic descriptions of actions elicit comparable activations across premotor and parietal regions (Behrendt et al., 2021; Boulenger et al., 2006; Sobierajewicz et al., 2016). Thus, despite the laboratory constraints, the present setup provides an ecologically valid window into real-world neural decision processes.

Second, the dataset size remains modest, involving twelve participants as an initial proof-of-concept. While sufficient for model convergence and statistical reliability, larger and more demographically diverse samples will be required to assess inter-individual variability and generalizability. Future work will focus on expanding the participant pool to include diverse age groups, cognitive profiles, and situational contexts (such as urgency, vigilance, and distraction) to examine how the human brain dynamically reconfigures cognitive resources under varying environmental demands.

Furthermore, scaling the current approach to personalized adaptation will require extending the model toward subject-specific calibration and cross-subject generalization. Transfer learning, federated learning, and continual learning paradigms offer promising directions to achieve such adaptability while preserving privacy and computational efficiency.

Finally, future work will integrate real-time human–robot interaction experiments, where decoded neural intentions can modulate robotic control policies directly. This extension will enable the closed-loop validation of the proposed predictive framework, advancing toward truly cognition-aware robotic systems that anticipate and align with human intentions in shared physical spaces.

## Conflicts of Interest

None

## Acknowledgments

The work presented in this paper was supported financially by the United States Department of Transportation, Center for Connected and Automated Transportation (CCAT) via Award# 69A3552348305 and by the United States National Science Foundation (NSF) via Award# SCC-IRG 2124857. The support of the CCAT and NSF is gratefully acknowledged. Any opinions and findings in this paper are those of the authors and do not necessarily represent those of the CCAT or the NSF.